\documentclass[aps,prc,twocolumn,superscriptaddress]{revtex4-2}
\usepackage{graphicx}
\usepackage{makecell}
\usepackage{color}
\usepackage[dvipsnames]{xcolor}
\usepackage[colorlinks=true,urlcolor=Blue,linkcolor=Blue]{hyperref}
\usepackage[all]{hypcap}
\usepackage{amsmath}
\newcommand{\YAL}{\textcolor{blue}}

\begin{document}



\title{$B\rho$-defined Isochronous Mass Spectrometry and Mass Measurements of $^{58}$Ni Fragments }

\author{M.~Zhang}
	\affiliation{CAS Key Laboratory of High Precision Nuclear Spectroscopy,  Institute of Modern Physics, Chinese Academy of Sciences, Lanzhou 730000, China}
	\affiliation{School of Nuclear Science and Technology, University of Chinese Academy of Sciences, Beijing 100049, China}
\author{X.~Zhou}
	\affiliation{CAS Key Laboratory of High Precision Nuclear Spectroscopy,  Institute of Modern Physics, Chinese Academy of Sciences, Lanzhou 730000, China}
	\affiliation{School of Nuclear Science and Technology, University of Chinese Academy of Sciences, Beijing 100049, China}
\author{M.~Wang}
	\email{wangm@impcas.ac.cn}
	\affiliation{CAS Key Laboratory of High Precision Nuclear Spectroscopy,   Institute of Modern Physics, Chinese Academy of Sciences, Lanzhou 730000, China}
	\affiliation{School of Nuclear Science and Technology, University of Chinese Academy of Sciences, Beijing 100049, China}
\author{Y.~H.~Zhang}
	\email{yhzhang@impcas.ac.cn}
	\affiliation{CAS Key Laboratory of High Precision Nuclear Spectroscopy,   Institute of Modern Physics, Chinese Academy of Sciences, Lanzhou 730000, China}
	\affiliation{School of Nuclear Science and Technology, University of Chinese Academy of Sciences, Beijing 100049, China}	
\author{Yu.~A.~Litvinov}
	\email{y.litvinov@gsi.de}
	\affiliation{CAS Key Laboratory of High Precision Nuclear Spectroscopy,   Institute of Modern Physics, Chinese Academy of Sciences, Lanzhou 730000, China}
	\affiliation{GSI Helmholtzzentrum f{\"u}r Schwerionenforschung, Planckstra{\ss}e 1, 64291 Darmstadt, Germany}	
\author{H.~S.~Xu}
	\affiliation{CAS Key Laboratory of High Precision Nuclear Spectroscopy,   Institute of Modern Physics, Chinese Academy of Sciences, Lanzhou 730000, China}
	\affiliation{School of Nuclear Science and Technology, University of Chinese Academy of Sciences, Beijing 100049, China}
\author{R.~J.~Chen}
	\affiliation{CAS Key Laboratory of High Precision Nuclear Spectroscopy,   Institute of Modern Physics, Chinese Academy of Sciences, Lanzhou 730000, China}
	\affiliation{GSI Helmholtzzentrum f{\"u}r Schwerionenforschung, Planckstra{\ss}e 1, 64291 Darmstadt, Germany}
\author{H.~Y.~Deng}
	\affiliation{CAS Key Laboratory of High Precision Nuclear Spectroscopy,   Institute of Modern Physics, Chinese Academy of Sciences, Lanzhou 730000, China}
	\affiliation{School of Nuclear Science and Technology, University of Chinese Academy of Sciences, Beijing 100049, China}
\author{C.~Y.~Fu}
	\affiliation{CAS Key Laboratory of High Precision Nuclear Spectroscopy,   Institute of Modern Physics, Chinese Academy of Sciences, Lanzhou 730000, China}
\author{W. W. Ge}
	\affiliation{CAS Key Laboratory of High Precision Nuclear Spectroscopy,   Institute of Modern Physics, Chinese Academy of Sciences, Lanzhou 730000, China}
\author{H.~F.~Li}
	\affiliation{CAS Key Laboratory of High Precision Nuclear Spectroscopy,   Institute of Modern Physics, Chinese Academy of Sciences, Lanzhou 730000, China}
	\affiliation{School of Nuclear Science and Technology, University of Chinese Academy of Sciences, Beijing 100049, China}
\author{T.~Liao}
	\affiliation{CAS Key Laboratory of High Precision Nuclear Spectroscopy,   Institute of Modern Physics, Chinese Academy of Sciences, Lanzhou 730000, China}
	\affiliation{School of Nuclear Science and Technology, University of Chinese Academy of Sciences, Beijing 100049, China}
\author{S.~A.~Litvinov}
	\affiliation{GSI Helmholtzzentrum f{\"u}r Schwerionenforschung, Planckstra{\ss}e 1, 64291 Darmstadt, Germany}
	\affiliation{CAS Key Laboratory of High Precision Nuclear Spectroscopy,   Institute of Modern Physics, Chinese Academy of Sciences, Lanzhou 730000, China}
\author{P.~Shuai}
	\affiliation{CAS Key Laboratory of High Precision Nuclear Spectroscopy,   Institute of Modern Physics, Chinese Academy of Sciences, Lanzhou 730000, China}
\author{J.~Y.~Shi}
	\affiliation{CAS Key Laboratory of High Precision Nuclear Spectroscopy,   Institute of Modern Physics, Chinese Academy of Sciences, Lanzhou 730000, China}
	\affiliation{School of Nuclear Science and Technology, University of Chinese Academy of Sciences, Beijing 100049, China}
\author{R.~S.~Sidhu}
	\affiliation{GSI Helmholtzzentrum f{\"u}r Schwerionenforschung, Planckstra{\ss}e 1, 64291 Darmstadt, Germany}
\author{Y. N. Song}
	\affiliation{CAS Key Laboratory of High Precision Nuclear Spectroscopy,   Institute of Modern Physics, Chinese Academy of Sciences, Lanzhou 730000, China}
	\affiliation{School of Nuclear Science and Technology, University of Chinese Academy of Sciences, Beijing 100049, China}
\author{M.~Z.~Sun}
	\affiliation{CAS Key Laboratory of High Precision Nuclear Spectroscopy,   Institute of Modern Physics, Chinese Academy of Sciences, Lanzhou 730000, China}
\author{S.~Suzuki}
	\affiliation{CAS Key Laboratory of High Precision Nuclear Spectroscopy,   Institute of Modern Physics, Chinese Academy of Sciences, Lanzhou 730000, China}
\author{Q.~Wang}
	\affiliation{CAS Key Laboratory of High Precision Nuclear Spectroscopy,   Institute of Modern Physics, Chinese Academy of Sciences, Lanzhou 730000, China}
	\affiliation{School of Nuclear Science and Technology, University of Chinese Academy of Sciences, Beijing 100049, China}
\author{Y.~M.~Xing}
	\affiliation{CAS Key Laboratory of High Precision Nuclear Spectroscopy,   Institute of Modern Physics, Chinese Academy of Sciences, Lanzhou 730000, China}
\author{X.~Xu}
	\affiliation{CAS Key Laboratory of High Precision Nuclear Spectroscopy,   Institute of Modern Physics, Chinese Academy of Sciences, Lanzhou 730000, China}
\author{T.~Yamaguchi}
	\affiliation{Department of Physics, Saitama University, Saitama 338-8570, Japan}
\author{X.~L.~Yan}
	\affiliation{CAS Key Laboratory of High Precision Nuclear Spectroscopy,   Institute of Modern Physics, Chinese Academy of Sciences, Lanzhou 730000, China}
\author{J.~C.~Yang}
	\affiliation{CAS Key Laboratory of High Precision Nuclear Spectroscopy,   Institute of Modern Physics, Chinese Academy of Sciences, Lanzhou 730000, China}
	\affiliation{School of Nuclear Science and Technology, University of Chinese Academy of Sciences, Beijing 100049, China}
\author{Y.~J.~Yuan}
	\affiliation{CAS Key Laboratory of High Precision Nuclear Spectroscopy,   Institute of Modern Physics, Chinese Academy of Sciences, Lanzhou 730000, China}
	\affiliation{School of Nuclear Science and Technology, University of Chinese Academy of Sciences, Beijing 100049, China}
\author{Q.~Zeng}
	\affiliation{School of Nuclear Science and Engineering, East China University of Technology, Nanchang 330013, China}
\author{X.~H.~Zhou}
	\affiliation{CAS Key Laboratory of High Precision Nuclear Spectroscopy,   Institute of Modern Physics, Chinese Academy of Sciences, Lanzhou 730000, China}
	\affiliation{School of Nuclear Science and Technology, University of Chinese Academy of Sciences, Beijing 100049, China}

\date{\today}

\begin{abstract}
A novel isochronous mass spectrometry, termed as $B\rho$-defined IMS, is established at the experimental cooler-storage ring CSRe in Lanzhou.
Its potential has been studied through high precision mass measurements of $^{58}$Ni projectile fragments.
Two time-of-flight detectors were installed in one of the straight sections of CSRe, thus enabling simultaneous measurements of the velocity and the revolution time of each stored short-lived ion.
This allows for calculating the magnetic rigidity $B\rho$ and the orbit length $C$ of each ion.
The accurate $B\rho(C)$ function has been constructed, which is a universal calibration curve used to deduce the masses of the stored nuclides.
The sensitivity to single stored ions, quickness, and background-free characteristics of the method 
are ideally suited to address nuclides with very short lifetimes and tiniest production yields.
In the limiting case of just a single particle, the attained mass resolving power allows one us to determine its mass-over-charge ratio $m/q$ with
a remarkable precision of merely $\sim5$~keV. Masses of $T_z=-3/2$ $fp$-shell nuclides are re-determined with high accuracy, and the validity of the isospin multiplet mass equation is tested up to the heaviest isospin quartet with $A=55$. The new masses are also used to investigate the mirror symmetry of empirical residual proton-neutron interactions. 
\end{abstract}


\maketitle

\section{Introduction}
Masses of atomic nuclei are needed in the study of a plethora of phenomena in nuclear structure and nuclear astrophysics, as well as for testing fundamental interactions and symmetries~\cite{Lunney2003,Blaum2006,ERONEN2016259,Dilling2018}. 
Masses of about 2550 nuclides are known experimentally today~\cite{Huang2021} while about 7000 nuclides are expected to exist~\cite{Erler-2012}.
Of present interest are the masses of nuclides lying far away from the valley of $\beta$-stability line~\cite{YAMAGUCHI2021}.
Especially masses of very neutron-rich nuclides are needed for modelling the r-process nucleosynthesis~\cite{Cowan2021}.
Such nuclides are inevitably short-lived and have tiny production yields, making their mass measurements extremely challenging.
Hence, measurement techniques capable of determining precisely the mass from a single, short-lived particle are especially demanded.

The isochronous mass spectrometry (IMS) at heavy-ion storage rings is an efficient and fast experimental technique~\cite{Hausmann2000,STADLMANN2004} well suited for mass measurements of exotic nuclei with short lifetimes down to several tens of microseconds. Since the pioneering experiments conducted at the ESR in GSI, Darmstadt~\cite{Hausmann2000,STADLMANN2004}, IMS has been established at the experimental cooler storage ring (CSRe) in IMP, Lanzhou, and 
at the Rare RI Ring (R3) in RIKEN, Saitama~\cite{Zhang2016,PhysRevLett.128.152701}.

To conduct the isochronous mass measurements, the beam energy has to be specially tuned in order to fulfill the isochronous condition. 
Thereby, the high mass resolving power can be achieved for the nuclides of interest~\cite{Hausmann2000}.
Usually, the isochronous condition is fulfilled only for the ion species in a limited range of mass-to-charge ($m/q$) ratios, while the resolving powers are inevitably deteriorated for the majority of ion species not fulfilling the isochronous condition. 
In addition, the momentum distributions of different ion species injected into the ring have different shapes,
often asymmetric due to production nuclear reaction mechanisms and limited acceptance of the transfer line and storage ring. 
This leads not only to large statistical uncertainties, but may also cause systematic deviations in the mass determination.

In order to improve the mass resolving power as well as to reduce the systematic deviation in a broad $m/q$ range, the magnetic-rigidity tagging ($B\rho$-tagging) IMS was realized at the FRS-ESR facility at GSI~\cite{Geissel_2005,Geissel2006,SUN20081} by inserting metal slits at the second dispersive focal plane of FRS~\cite{GEISSEL1992}. 
In CSRe, an in-ring slit was used to restrict the magnetic-rigidity ($B\rho$) acceptance of stored fragments~\cite{LIU2020}. By using these two techniques, the mass resolving powers have been improved significantly and the systematic deviations have been largely reduced~\cite{Geissel_2005,Geissel2006,SUN20081,Xu2016,Zhang2018,Xing-2018,Xu2019}. 
However, both approaches have a dramatic drawback of losing transmission efficiency, which is not tolerable for the mass measurements of exotic nuclei with very low production yields. Therefore, a novel idea was conceived to measure the velocity of each stored ion~\cite{Geissel_2005,Geissel2006,SUN20081,WALKER2013}. 
Recently we have installed two identical TOF detectors 18 m apart in one of the straight sections of CSRe~\cite{Xing2015}. 
This enabled simultaneous precision measurements of revolution time $T$ as well as velocity $v$ of each stored
ion. Using these two parameters, a novel mass measurement technique, {\it $B\rho$-defined IMS}, has been
established, which--in a limiting case of just a single
event--is capable of determining the mass-to-charge ratio,
$m/q$, with an unprecedented precision of about 5 keV/$q$.

In this paper, the new scheme of IMS using the two measured quantities, $T$’s and $v$’s, is described. 
Experiment and data processing
are briefly introduced in section~\ref{sec:experiment}. 
We describe the details of the $B\rho$-defined IMS in section~\ref{sec:brims}, focussing on the construction of the $B\rho=B\rho(C)$ function and the mass determination. 
In Section~\ref{sec:mass}, the re-determined masses for some $fp$-shell nuclides are reported and used to investigate the mirror symmetry of empirical residual proton-neutron ($pn$) interactions and to test the validity of the isospin multiplet mass equation. The conclusion and outlook are given in Section~\ref{sec:conclusion}.

\section{Experiment and revolution time spectrum}\label{sec:experiment}

\begin{figure}[htb]
	\centering
	\includegraphics[scale=1.0]{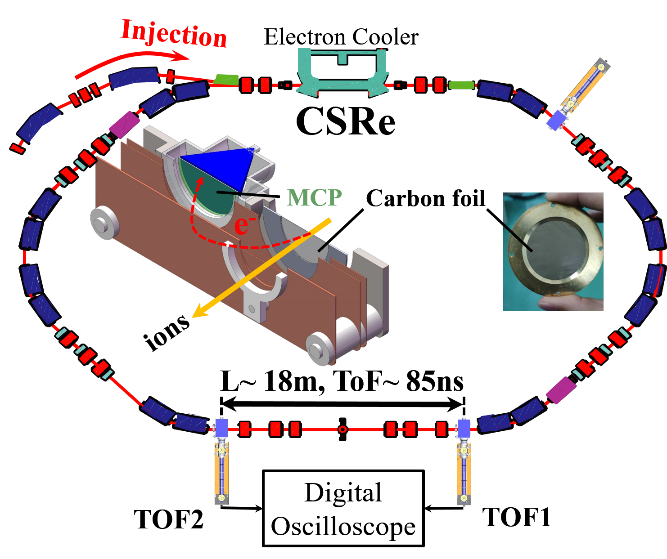}
	\caption{Schematic view of the CSRe with the arrangement of the TOF detectors.}
	\label{fig:csre}
\end{figure}

The nuclides of interest were produced by fragmenting 440 MeV/u $^{58}$Ni$^{19+}$ primary projectiles on 15~mm thick $^9$Be target. They were selected with the in-flight fragment separator RIBLL2~\cite{XIA2002,ZHAN2010}. Every 25 seconds, a cocktail beam including the nuclides of interest was injected into and stored in CSRe. The duration of each measurement was merely 400~$\rm{\mu s}$. 
Figure~\ref{fig:csre} presents the schematic view of the storage ring CSRe having a circumference of 128.8~m on the central orbit. 
CSRe was tuned into the isochronous mode with expected $\gamma_t=1.365$. The whole $B\rho$-acceptance was about $\pm0.2\%$. The RIBLL2-CSRe system was set to a fixed central magnetic rigidity of $B\rho=5.471$~Tm. At the employed relativisitc energies, the produced fragments were fully stripped of bound electrons. The primary beam energy was selected according to the LISE++ simulations~\cite{TARASOV20084657} 
such that the $^{55}$Cu$^{29+}$ ions had the most probable velocity corresponding to $\gamma \sim \gamma_t$.

Two identically designed TOF detectors were installed 18~m apart in the straight section of CSRe (see Fig.~\ref{fig:csre}). 
Each detector consists of a thin carbon foil ($\phi$40~mm, 18~$\rm\mu g/cm^2$) and a set of micro-channel plates (MCP)~\cite{ZHANG20141}. 
When an ion passed through the carbon foil, secondary electrons were released from the foil surface and isochronously guided to MCP 
by perpendicularly arranged electric and magnetic fields. 
Fast timing signals from the two MCPs were recorded by a digital oscilloscope at a sampling rate of 50~GHz. 
The time resolutions of the TOF detectors were determined off-line to be $\sigma=20$ - 40~ps by using $^{241}$Am $\alpha$ source. 
Details on the detector performance can be found in Refs.~\cite{ZHANG20141,ZHANG201438}.

\begin{figure*}[htb]
	\centering
	\includegraphics[scale=0.85]{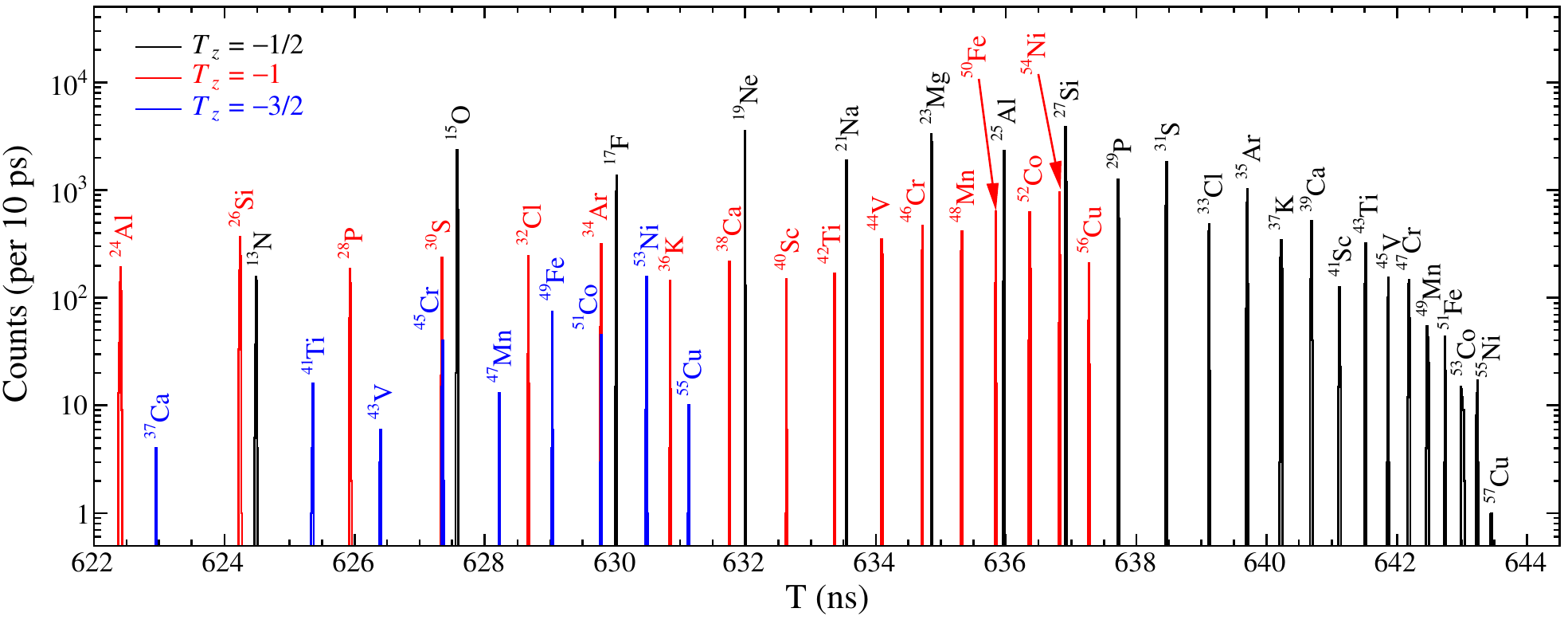}
	\caption{Revolution-time spectrum extracted from the raw data acquired by the two time-of-flight detectors. 
		Unambiguous particle identification has been made as described in Ref.~\cite{XING2019}.
		Different colours indicate the series of nuclides with a constant isospin projection $T_z=(N-Z)/2$, see legend.}
	\label{fig:tspec}
\end{figure*}

For each particle circulating in the ring, two time sequences, 
i.e. the time stamps when passing the two TOF detectors, $t_{\rm TOF1}(N)$ and $t_{\rm TOF2}(N)$, 
as a function of the revolution number, $N$, were extracted from the recorded signals. 
Particles stored for more than 230~$\rm{\mu s}$ in the ring were used in the data analysis.
The revolution times, $T_{\rm TOF1}$ and $T_{\rm TOF2}$, of each stored ion 
were deduced independently following the procedures described in Ref.~\cite{XING2019}. 
We define $(T_{\rm TOF1}+T_{\rm TOF2} )/2$ at the middle revolution number to be the revolution time. 
All individual revolution times were put into a histogram forming an integrated time spectrum as shown in Fig.~\ref{fig:tspec}. 
Particle identification was made following the procedures described in Ref.~\cite{XING2019}. Three series of nuclides with $T_z=-1/2,-1$, and $-3/2$ 
can be seen
in Fig.~\ref{fig:tspec}. 

\begin{figure}[htb]
	\centering
    \includegraphics[scale=0.4]{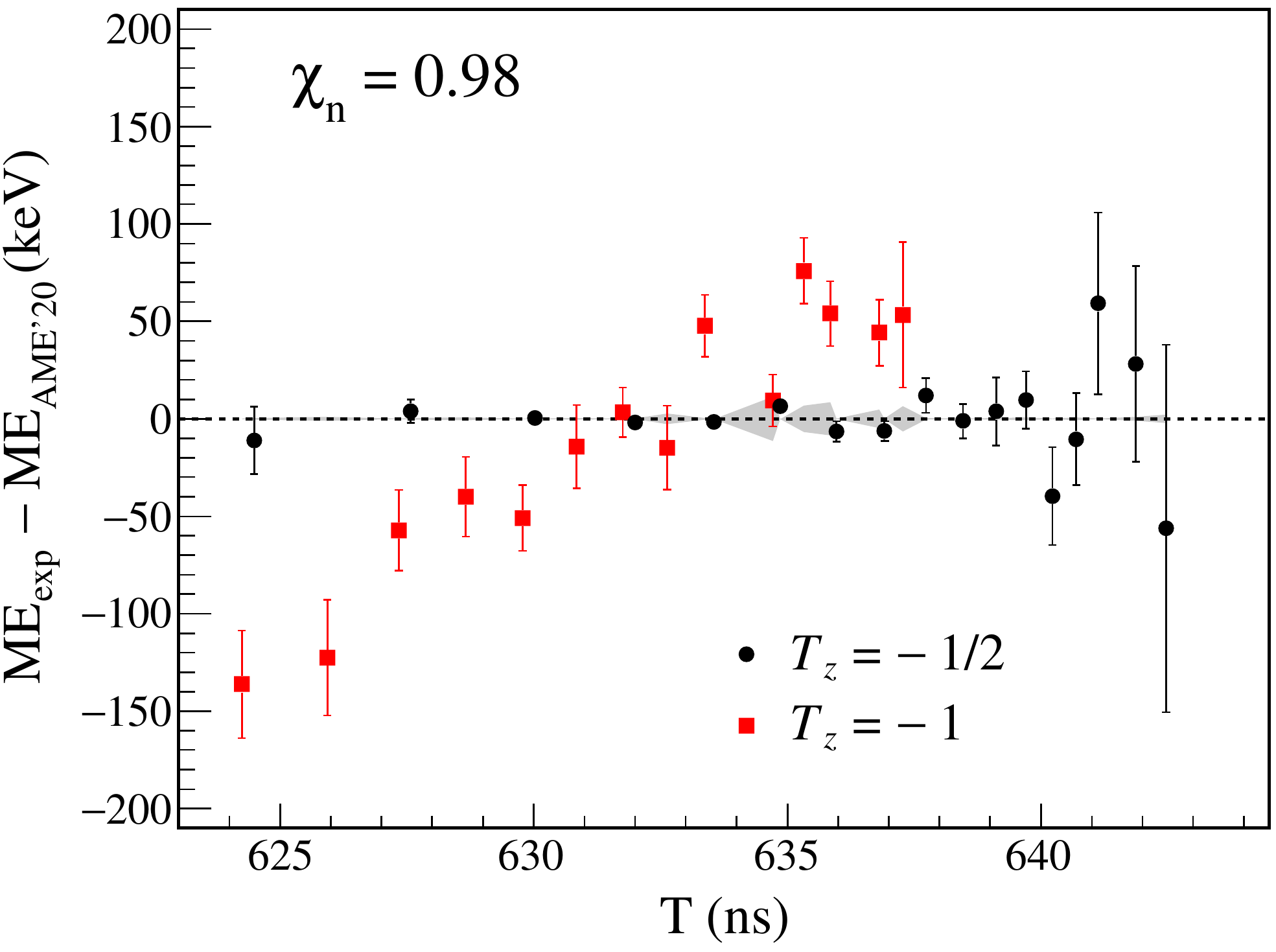}	
	\caption{Comparison of re-determined mass excesses of $T_z=-1$ nuclides (open red squares) with literature values~\cite{Wang2021} using the $T_z=-1/2$ nuclides (filled black circles) as calibrants. Masses are determined from the revolution time spectrum following the procedures described in~\cite{Zhang2018}.	
	}
	\label{fig:NoCor}
\end{figure}

Nuclear masses can be deduced from the revolution time spectrum, 
see e.g. our previous work~\cite{Zhang2018}. 
Using the $T_z=-1/2$ nuclides to calibrate the spectrum, the re-determined masses of the $T_z=-1$ nuclides are compared with the well-known literature values in Fig.~\ref{fig:NoCor}.
One observes systematic deviations for the $T_z=-1$ nuclides. 
Such systematic deviations are caused mainly by different momentum distributions and energy losses of the two series of nuclides with $T_z=-1/2$ and $-1$. 
This phenomenon was also observed in our previous experiments, and could be reduced by limiting the $B\rho$ acceptance of the ring at a cost of a dramatic loss of the transmission efficiency. 
In the following, we will show that, by using the velocity information of stored ions, not only can the systematic deviations be removed, but also the mass resolving powers are improved significantly over a wide range of revolution times.
\section{$B\rho$-defined IMS}\label{sec:brims}
\subsection{Principle}
The revolution time, $T$, of an ion circulating in a storage ring is given by 
\begin{equation}\label{eq:t}
T=\frac{C}{v},
\end{equation}
with $C$ and $v$ being the orbit length and velocity of the ion, respectively. 
Magnetic rigidity, $B\rho$, is defined as
\begin{equation}\label{eq:brho}
B\rho=\frac{m}{q}\gamma v,
\end{equation}
where the Lorentz factor $\gamma=1/\sqrt{1-\beta^2}$ with $\beta$ being the velocity in units of the speed of light in vacuum $v_c$.
Stored ions with identical $B\rho$ values move on the same mean orbits in the storage ring.
By combining Eqs.~(\ref{eq:t}) and~(\ref{eq:brho}), one obtains
\begin{equation}\label{eq:mvq}
\frac{m}{q}=\frac{B\rho}{\gamma v}=B\rho \sqrt{\Bigg(\frac{T}{C}\Bigg)^2-\Bigg(\frac{1}{v_c}\Bigg)^2}.
\end{equation}
Equation~(\ref{eq:mvq}) is the basic formula for any kind of $B\rho$-Time-of-Flight mass spectrometry~\cite{YAMAGUCHI2021103882}. 
According to this equation, $m/q$ values can be determined if the quantities $(B\rho,v)$ or equivalently $(B\rho,T,C)$ are measured with high precision. 
However, it is difficult to precisely measure the absolute $B\rho$ values. In the following, a method is decribed to obtain the relation of $B\rho$ versus $C$ using the directly measured quantities in the experiment.

\subsection{Functional form of the $B\rho(C)$ curve} \label{ssec:brhocfunc}

If the transition point of the storage ring, $\gamma_t$, defined as~\cite{Hausmann2000} 
\begin{equation}\label{eq:gt}
\frac{1}{\gamma_t^2}=\frac{\Delta C /C}{\Delta B\rho /B\rho},
\end{equation}
is constant within the entire $B\rho$-acceptance of the ring, the $B\rho(C)$ function should have a simple form
\begin{equation}\label{eq:fcfunc0}
B\rho(C)=B\rho_0\left(\frac{C}{C_0}\right)^{\gamma_t^2},
\end{equation}
with $B\rho_0$ and $C_0$ being the reference parameters. 
In reality, the experimental $B\rho_{\rm exp}$ vs $C_{\rm exp}$ relation cannot be well described by Eq.~(\ref{eq:fcfunc0}) 
because $\gamma_t$ is not constant for all orbit lengths~\cite{Chen2018,Ge2018,ZHANG2022166329}. 
Thus Eq.~(\ref{eq:fcfunc0}) was extended by adding arbitrarily an extra term
\begin{equation}\label{eq:fcfunc}
B\rho(C)=B\rho_0\left(\frac{C}{C_0}\right)^{K}+a_1e^{-a_2(C-C_0)}.
\end{equation}
The free parameters $B\rho_0$, $C_0$, $a_1$, $a_2$, and $K$ can be determined via fitting the experimental data. 
We emphasize that Eq.~(\ref{eq:fcfunc}) characterizes the ion's motion in the ring, i.e., all experimental $B\rho _{\rm exp}$ and $C_{\rm exp}$ data should fall onto an identical curve in the $B\rho$ - $C$ plane which can be described by the $B\rho(C)$ function. 
The second term of Eq.~(\ref{eq:fcfunc}) is optionally chosen to describe the effect of the variable $\gamma_t$, and its expression depends on the specific beam-optical setting of the ring.

\subsection{Determination of the $B\rho(C)$ function }\label{sec:determination}

$B\rho(C)$ function can be determined using the measured quantities of the known-mass nuclides. In the present experiment, on average $\sim$15 ions were stored in one injection in CSRe.
Each of them is easily identified in the recorded time stamp sequences.
The velocity of each ion was determined according to 
\begin{equation}\label{eq:vel}
v=\frac{L}{\Delta t_{\rm TOF}}=\frac{L}{\Delta t_{\rm exp}-\Delta t_{\rm d}},
\end{equation}
where $L$ is the distance between the two TOF detectors, 
$\Delta t_{\rm TOF}=\Delta t_{\rm exp}-\Delta t_{\rm d}$ 
is the flight time of the particle from TOF1 to TOF2 at the middle revolution number, and 
$\Delta t_{\rm d}$ is the time delay difference of the timing signals from the two detectors. 

The $\Delta t_{\rm exp}$ values were determined from the time sequences measured by the two TOF detectors~\cite{Zhou2021} 
with the mean uncertainties of 2.0 - 6.4 ps for different ion species corresponding to the relative uncertainties of (2.2 - 7.2)$\times10^{-5}$. 
$L$ and $\Delta t_{\rm d}$ were measured offline by using a dedicated laser setup~\cite{Yan2019}.
In this experiment, $L$ and $\Delta t_{\rm d}$ were determined from experimental data to be $L=18.046$~m and $\Delta t_{\rm d}=-146.83$~ps, see Appendix~\ref{app:pars}.

\begin{figure}[htbp]
	\centering
	\includegraphics[scale=0.40]{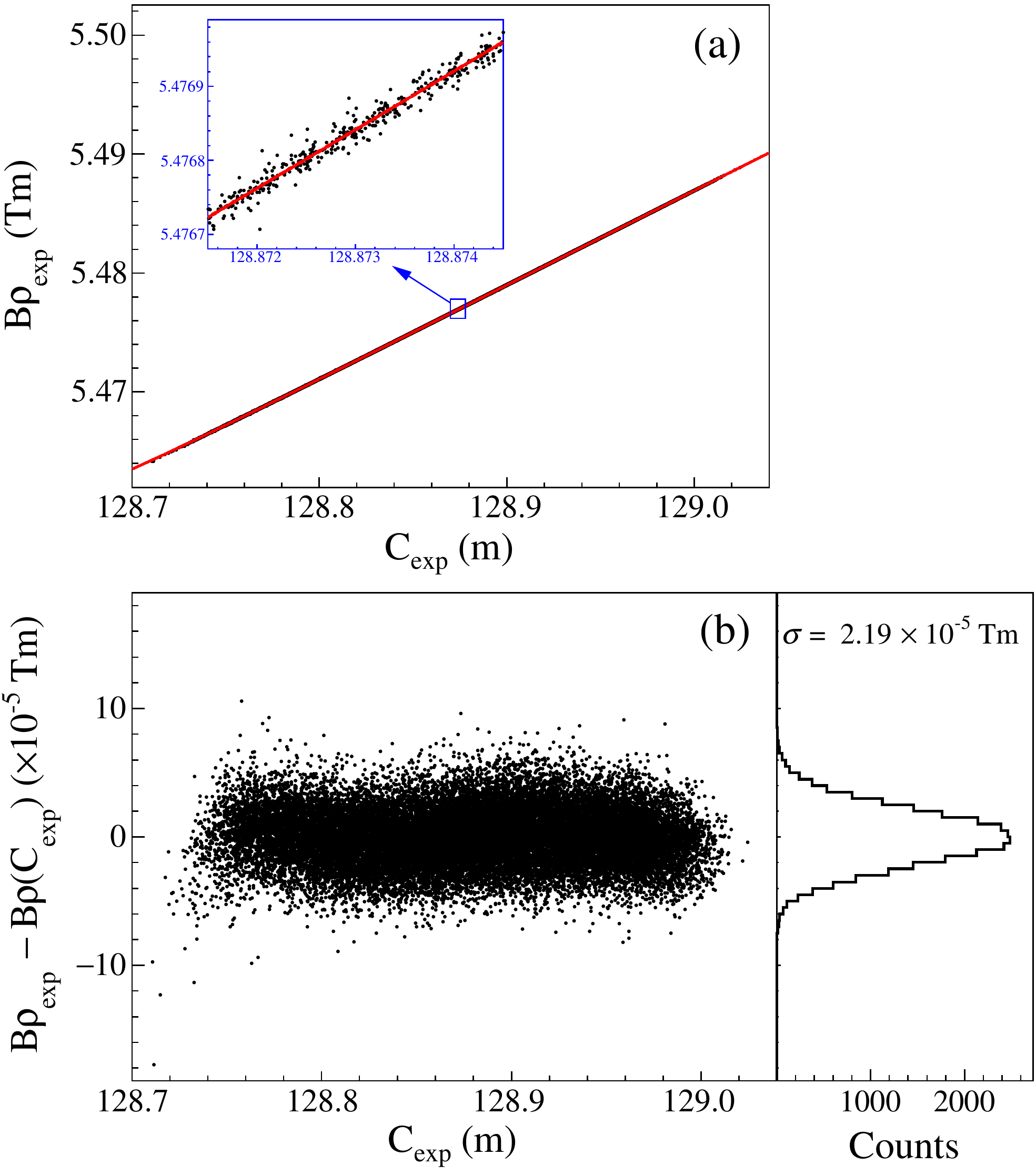}
	\caption{(a) Plot of $B\rho_{\rm exp}$ versus $C_{\rm exp}$ with the fit results (solid line) according to Eq.~(\ref{eq:fcfunc}). 
		(b) Scatter plot of the fit residuals, $R_{B\rho}= B\rho_{\rm exp}-B\rho(C_{\rm exp})$, as a function of $C_{\rm exp}$ and its projected spectrum. Only a part of the injections were used.}
	\label{fig:gbrhoc}
\end{figure}

The ion species with well-known masses (mass uncertainties less than 5~keV) and having more than 100 events were used to establish the $B\rho(C)$ function.
Using the redetermined values of $L$ and  $\Delta t_{\rm d}$, the velocities for all stored ions were obtained.
The orbit length of the $i^{\rm th}$ ion was deduced via $C^i_{\rm {exp}}=v^i_{\rm {exp}} T^i_{\rm{exp}}$.
The corresponding magnetic rigidity was calculated with Eq.~(\ref{eq:brho}).
A part of the obtained dataset $(B\rho_{\rm{exp}}^i,C_{\rm{exp}}^i)$, where the magnetic fields of CSRe were relatively stable, is plotted in Fig.~\ref{fig:gbrhoc}(a). 

The fit according to Eq.~(\ref{eq:fcfunc}) is shown with red solid line in Fig.~\ref{fig:gbrhoc}(a). The fit residuals, $R_{B\rho}= B\rho_{\rm exp}-B\rho(C_{\rm exp})$, 
exhibit a Gaussian-like distribution with a mean $R_{B\rho}$ value at zero (see Fig.~\ref{fig:gbrhoc}(b)), 
indicating that Eq.~(\ref{eq:fcfunc}) is a good approximation of the expected $B\rho(C)$ function. Once the $B\rho (C)$ function is established, the mass-to-charge ratio of any stored ion can be directly obtained according to Eq.~(\ref{eq:mvq}). 

\subsection{Uncertainty of the \texorpdfstring{$B\rho(C)$}~ function}\label{ssec:emvq} 
\begin{figure}[htb]
	\centering
	\includegraphics[scale=0.4]{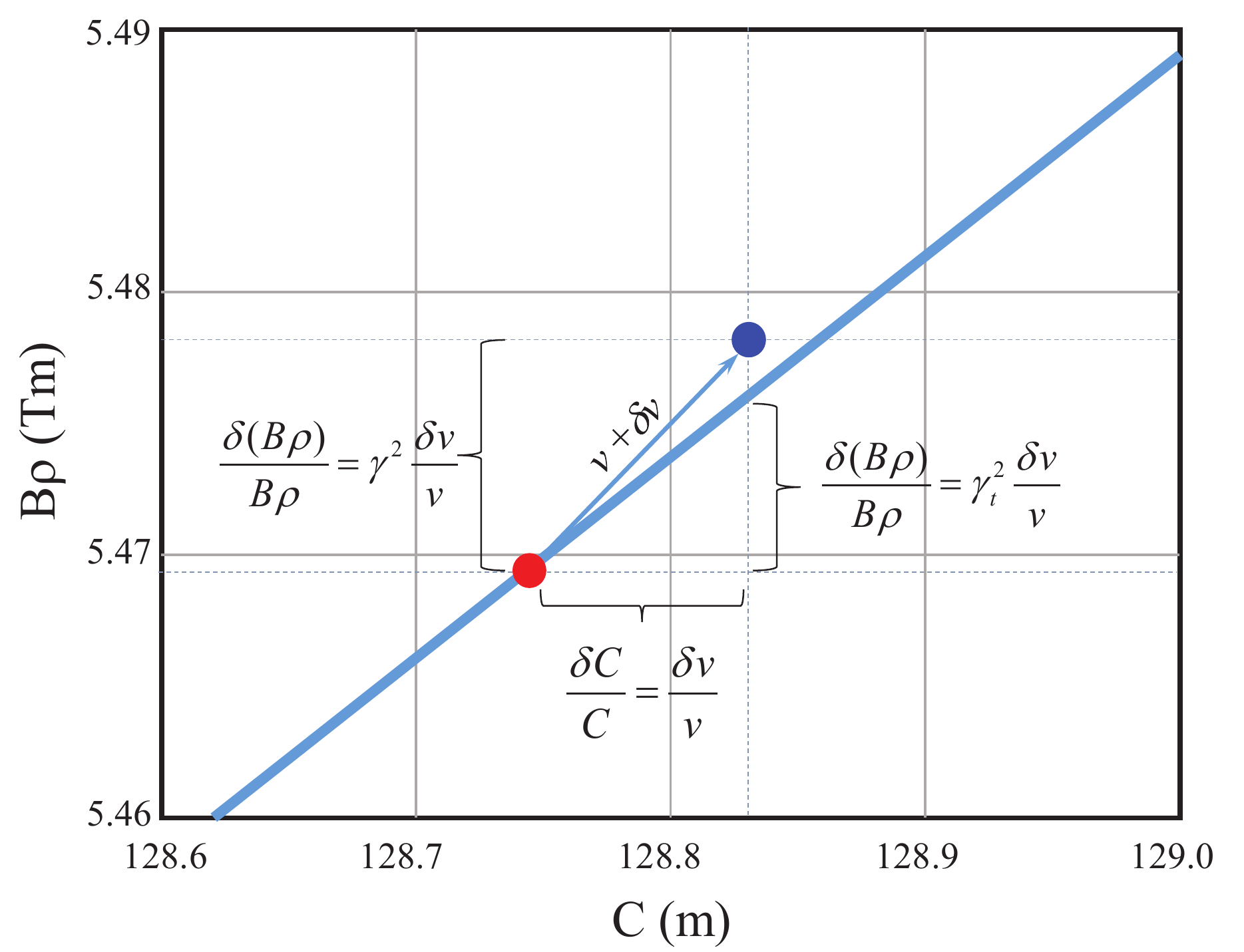}
	\caption{Schematic plot of the $B\rho(C)$ relation.
		The parameters for an example ion with a true velocity $v$ are indicated by the filled red circle.
		Due to the limited measurement accuracy, the velocity of this ion was determined as $v+\delta v$.
		The corresponding deviations of the magnetic rigidity $\delta B\rho/B\rho$ and orbit length $\delta C/C$ values are shown.
	}
	\label{fig:brhoerr}
\end{figure}

The measured velocities inevitably have uncertainties.
The $B\rho(C)$ curve is shown schematically in Fig.~\ref{fig:brhoerr}.

Let the measured velocity for a given ion deviate from the true value $v$ by $\delta v$.
Then the corresponding deviation $\delta B\rho$ from the true $B\rho$ value, $\delta B\rho/B\rho=\gamma^2\delta v/v$, 
and analogously $\delta C/C=\delta v/v$. 
The $\delta B\rho/B\rho$ can also be determined through $\delta C/C$ and the slope of the $B\rho(C)$ curve, see Fig.~\ref{fig:brhoerr}, 
giving $\delta B\rho/B\rho = \gamma_t^2\delta v/v$.
The two determined $\delta B\rho/B\rho$ values differ by $(\gamma^2-\gamma_t^2)\delta v/v$. 
This indicates a strong correlation between $B\rho$ and $C$ values.
Therefore, although the relative velocity uncertainty is in the order of $\sim 10^{-5}$, the error band of the $B\rho(C)$ curve is much smaller.

\begin{figure}[htb]
	\centering
	\includegraphics[scale=0.45]{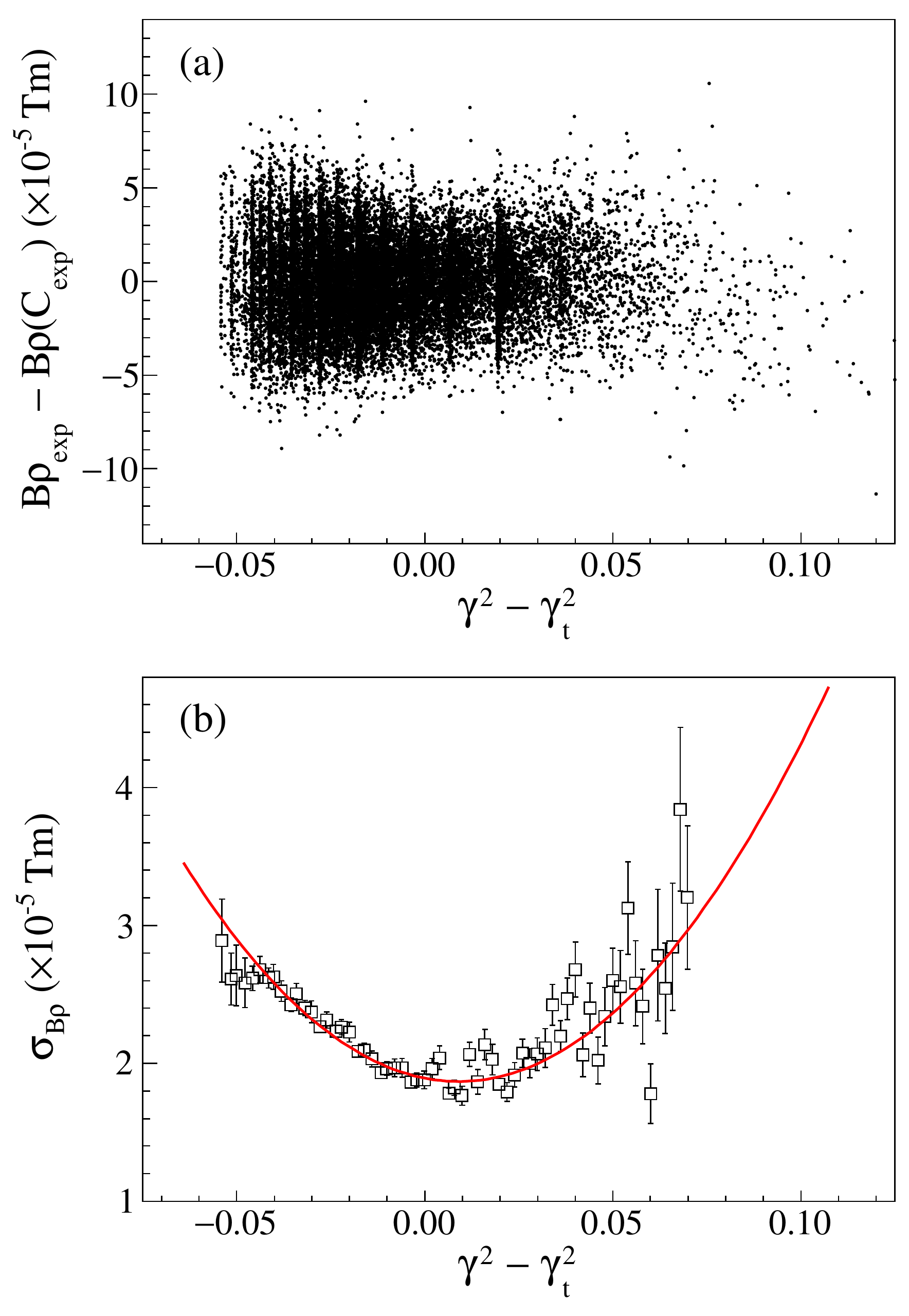}
	\caption{(a) Scatter plot of fit residuals, 
		$R_{B\rho}= B\rho_{\rm exp}-B\rho(C_{\rm exp})$ versus $\gamma^2-\gamma_t^2$.
		Only nuclides with mass uncertainties below 5~keV and with more than 100 detected ions are considered.
		(b) Variation of standard deviations $\sigma_{B\rho}(\gamma,\gamma_t)$ as a function of $\gamma^2-\gamma_t^2$. 
		The solid line is the fit result according to Eq.~(\ref{eq:sigmabr}). Only a part of the injections were used.}
	\label{fig:brhosig}
\end{figure}

The scatter plot of fit residuals, $R_{B\rho}= B\rho_{\rm exp}-B\rho(C_{\rm exp})$, 
as a function of the deviation from the isochronicity window, $\gamma^2-\gamma_t^2$, is presented in Fig.~\ref{fig:brhosig}(a). 
Determined standard deviations, denoted as $\sigma_{B\rho}(\gamma,\gamma_t)= \sigma(R_{B\rho})$, are shown in Fig.~\ref{fig:brhosig}(b). 
The variation of $\sigma_{B\rho}(\gamma,\gamma_t)$ versus $\gamma^2-\gamma_t^2$  is fitted by using a parabolic function
\begin{equation}\label{eq:sigmabr}
{\sigma_{B\rho}(\gamma,\gamma_t)}=b_0+b_1(\gamma^2-\gamma_t^2)+b_2(\gamma^2-\gamma_t^2)^2,
\end{equation}
where $\gamma_t$ is deduced from the $B\rho(C)$ function according to 
\begin{equation}\label{eq:gt2}
\gamma_t^2=\frac{dB\rho/B\rho}{dC/C}=\frac{C}{B\rho(C)}\frac{dB\rho(C)}{dC}.
\end{equation}

The fit result is shown with the solid line in Fig.~\ref{fig:brhosig}(b).
The obtained uncertainty $\sigma_{B\rho}(\gamma,\gamma_t)$ can be applied to all stored ions. 
Consequently, the relative uncertainty of each $(m/q)_{\rm exp}^i$ value, including the ion species with unknown masses, can be calculated
in the event-by-event analysis via the following expression
\begin{equation}\label{eq:sigmvq}
\frac{\sigma[(m/q)_{\rm exp}^i]}{(m/q)_{\rm exp}^i}=\frac{\sigma_{B\rho}(\gamma_{\rm exp}^i,\gamma_t)}{B\rho(C_{\rm exp}^i)},~~i=1,~2,~3,~....
\end{equation}

In principle, the resulting $B\rho(C)$ function can be directly employed to calculate the $m/q$ values according to Eq.~(\ref{eq:mvq}). However, in the present experiment, the magnetic fields of CSRe varied in time which led to large fluctuations of measured quantities. Therefore, the values such as $B\rho$ and $v$ derived directly from the experiment should be corrected in order to achieve high accuracy in the mass determination.

\subsection{Correction for the magnetic field drifts}

\begin{figure*}[htbp]
	\centering
	\includegraphics[scale=0.70]{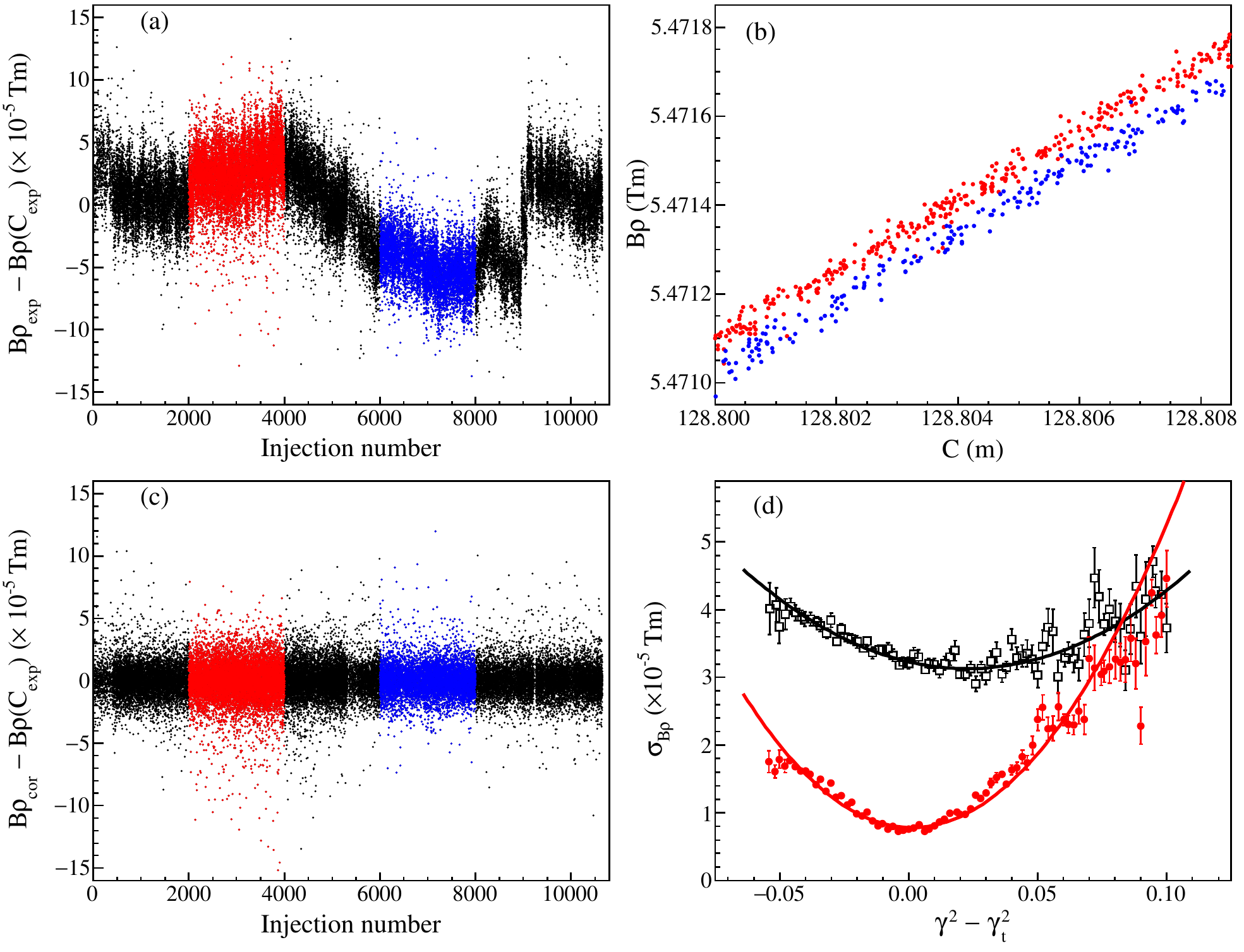}
	\caption{(a) Scatter plot of fit residuals, $R_{B\rho}^i= B\rho_{\rm{cor}}^i-B\rho(C_{\rm{exp}}^i)$, as a function of injection number 
		illustrating slow variations of the magnetic fields of the CSRe. 
		(b) $B\rho(C)$ plot for two magnetic field values. The points shown in red and blue colors correspond to the regions marked with the same color in the left panel.
		(c) Same as (a) but with the corrected $ B\rho_{\rm{cor}}$. 
		(d) Standard deviations, $\sigma_{B\rho}(\gamma,\gamma_t)$, as a function of $\gamma^2-\gamma_t^2$ before (black squares) and after (red circles) the correction procedure. The solid curves are the fits according to Eq.~(\ref{eq:sigmabr}).
	}
	\label{fig:rbrinj}
\end{figure*}

In a constant magnetic field $B_0$, the magnetic rigidities and orbit lengths of all ions must fall onto an identical $B\rho(C)$ curve.
However, the magnetic fields of CSRe vary in time which lead to strong deviations.
There were several methods developed for correction of the magnetic field drifts in the CSRe experiments~\cite{XING2019,Tu2011,Tu2011prl}.
In the present work, we adopt a much more precise correction method by using the data from the two TOF detectors.

The changes of the dipole magnetic fields lead to an up-down shifts of the $B\rho(C)$ curve. 
This is clearly seen in Fig.~\ref{fig:rbrinj}(a) which shows the fit residuals, $R_{B\rho}= B\rho_{\rm exp}-B\rho(C_{\rm exp})$, as a function of injection number.
To achieve the high mass resolving power, the effects due to magnetic field drifts are corrected as follows.

Let there be $N_s$ ions stored in an individual injection which is characterized by the magnetic field $B$.
The determined quantities $\{C_{\rm exp}^i,T_{\rm exp}^i,v_{\rm exp}^i,~i=1,2,...,N_s\}$ 
need to be corrected to $\{C_{\rm cor}^i,T_{\rm cor}^i,v_{\rm cor}^i,~i=1,2,...,N_s\}$ corresponding to a reference setting with $B_0$. 
To obtain the latter ones, $C_{\rm exp}^i=C_{\rm cor}^i$ should be confined (equivalent to a common radius $\rho_{\rm exp}^i=\rho_{\rm cor}^i$). Then 
the ratio of the magnetic fields, $M$, can be deduced from the experimental data according to
\begin{equation}\label{eq:ratiom}
M=\left(\frac{B}{B_0}\right)_{\rm CSRe}=\left(\frac{B\rho_{\rm{exp}}^i}{B\rho_{\rm{cor}}^i}\right)_{\rm ion},~~i=1,~2,~...,~N_s.
\end{equation}

For a specific injection, the ratio $M$ is constant for all $N_s$ stored ions. 
Hence, 
\begin{equation}\label{eq:brcor}
B\rho_{\rm{cor}}^i= \frac{1}{M}B\rho_{\rm{exp}}^i,~~i=1,~2,~...,~N_s,
\end{equation}
\begin{equation}\label{eq:vcor}
(\gamma v)_{\rm{cor}}^i=\frac{1}{M}(\gamma v)_{\rm{exp}}^i,~~i=1,~2,~...,~N_s.
\end{equation}
By using the measured velocities, $B\rho$ values can be precisely determined for nuclides with well-known masses.
For this purpose, nuclides with tabulated mass uncertainties less than 5~keV and recorded for more than 100 counts have been utilized.

\begin{figure*}[htbp]
	\centering
	\includegraphics[scale=0.85]{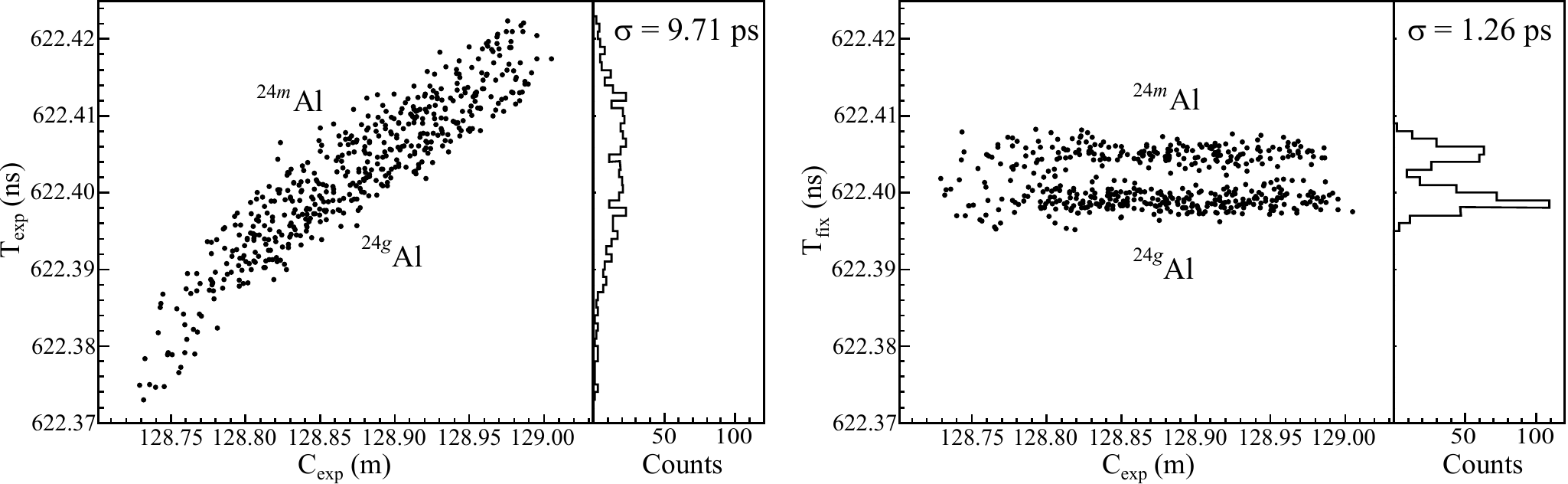}
	\caption{Scatter plots of $T_{\rm exp}$ and $T_{\rm fix}$ versus $C_{\rm exp}$ for $^{24}$Al ions. The excitation energy of the isomeric state $^{24m}$Al is 425.8(1)~keV~\cite{Huang2021}.}
	\label{fig:t24al}
\end{figure*}

The procedure was done iteratively.
First, the data $\left\{B\rho_{\rm{exp}},C_{\rm{exp}}\right\}$ were fitted with Eq.~(\ref{eq:fcfunc}) giving the initial $B\rho(C)$ function.
By analyzing the distribution of the fit residuals $B\rho_{\rm{exp}}-B\rho(C_{\rm{exp}})$, 
the estimated uncertainty $\sigma_{B\rho}$ was obtained for each ion as a function of $\gamma^2-\gamma_t^2$. 
Then, the expected $M$ value for an individual injection with $N_p$ known-mass nuclides (note $N_p\le N_s$) can be deduced from experimental data according to
\begin{equation}\label{eq:mexp}
M = \frac{1}{\sum_i^{N_p}{w_i}}{\sum_i^{N_p}{w_i\left[\frac{B\rho_{\rm{exp}}^i}{B\rho(C_{\rm{exp}}^i)}\right]}},~~i=1,~2,~...,~N_p,
\end{equation}
where
\begin{equation}\label{eq:weight}
\frac{1}{w_i} = \frac{[\sigma_{B\rho}(\gamma_{\rm exp}^i,\gamma_t^i)]^2}{[B\rho(C_{\rm exp}^i)]^2},~~i=1,~2,~...,~N_p.
\end{equation}
The obtained $M$ value provides the corrected $B\rho_{\rm{cor}}^i$ and $(\gamma v)_{\rm{cor}}^i$ via Eqs.~(\ref{eq:brcor}) and~(\ref{eq:vcor}). 
The corrected data $\{B\rho_{\rm{cor}}^i,C_{\rm{exp}}^i\}$ for the ions from all injections 
were used to establish a new $B\rho'(C)$ function and a new $\sigma'_{B\rho}$  estimation,
which were used in the next iteration.
The iterations were repeated until the convergence was reached.

Figure~\ref{fig:rbrinj}(c) shows the fit residuals after the correction procedure, $R_{B\rho}^i=B\rho_{\rm{cor}}^i-B\rho(C_{\rm{exp}}^i)$, 
as a function of the injection number.
The slow variations of the magnetic field drifts have been nearly completely removed.
Figure~\ref{fig:rbrinj}(d) illustrates the reduced standard deviations $\sigma_{B\rho}(\gamma,\gamma_t)$ as a function of $\gamma^2-\gamma_t^2$ obtained by analyzing all injections.

After the field-drift correction, the obtained values $\{B\rho_{\rm cor}^i,v_{\rm cor}^i\}$ were available for all ion species. Their $m/q$ values and the corresponding uncertainties were then calculated through Eqs.~(\ref{eq:mvq}) and~(\ref{eq:sigmvq}) using the final $B\rho(C)$ function.

\subsection{Improvements in respect to the conventional IMS}

In the conventional IMS, the masses of interest are determined directly from the measured revolution time spectrum. Using the $B\rho$-defined IMS, the $m/q$ values of nuclides can be determined in the event-by-event analysis according to Eq.~(\ref{eq:mvq}). To compare with the revolution time spectrum shown in Fig.~\ref{fig:tspec}, the individual $m/q$ values were transformed into a new revolution time spectrum, $T_{\rm fix}$, at a fixed magnetic rigidity, $B\rho_{\rm fix}=5.4758$ Tm, 
according to
\begin{equation}\label{eq:tfix}
T_{\rm fix}^i = C_{\rm fix}\sqrt{\frac{1}{B\rho_{\rm fix}^2}\left(\frac{m}{q}\right)_i^2 + \left(\frac{1}{v_c}\right)^2},~~i=1,~2,~3,~...
\end{equation}
The scatter plots of $T_{\rm exp}$ and $T_{\rm fix}$ versus $C_{\rm exp}$ for $^{24}{\rm Al}^{13+}$ ions are shown in Fig.~\ref{fig:t24al}. 
Note that the revolution time peak of $^{24}{\rm Al}^{13+}$ ions lies at the edge of the revolution time spectrum at about 622.5~ns, see Fig.~\ref{fig:tspec}.
It is obvious that the two states of $^{24}$Al, separated by the mass difference of 425.8(1)~keV~\cite{Huang2021}, cannot be resolved in the $T_{\rm exp}$ spectrum, 
while the two peaks can be clearly separated in the $T_{\rm fix}$ spectrum. 

\begin{figure}[htb]
	\centering
	\includegraphics[scale=0.40]{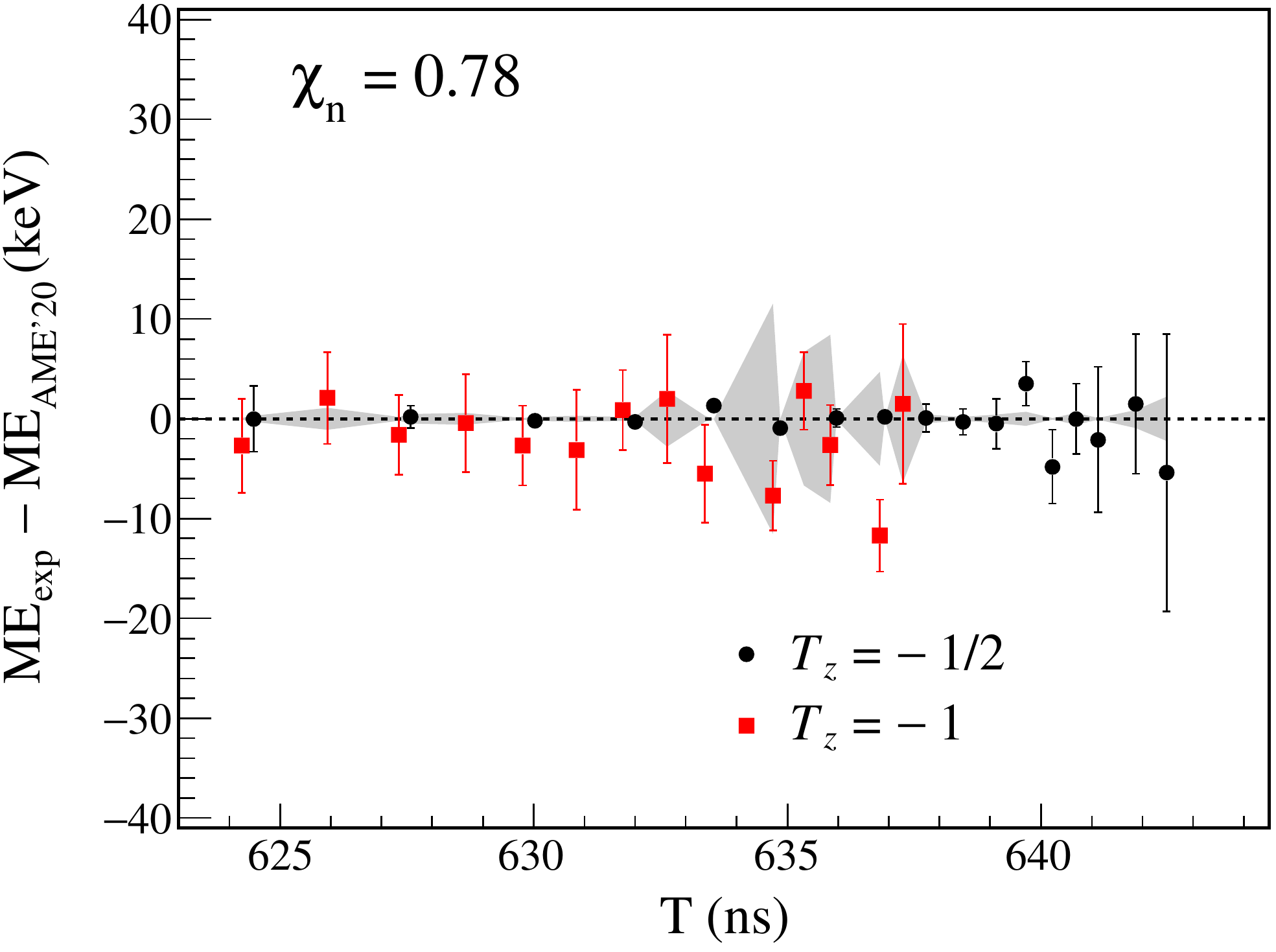}
	\caption{Comparison of re-determined mass excesses of $T_z=-1$ nuclides (open red squares) with literature values~\cite{Wang2021} using the $T_z=-1/2$ nuclides (filled black circles) as calibrants. Masses are determined with the $B\rho$-defined IMS.	
	}
	\label{fig:corrected}
\end{figure}

Figure~\ref{fig:corrected} shows a comparison of the re-determined masses of the $T_z=-1$ nuclides with the well-known literature values. 
Note that the $T_z=-1/2$ nuclides were used as calibrants in the analysis, and the masses of the $T_z=-1 $ nuclides were assumed to be unknown. 
One sees that not only the statistic uncertainties significantly decreased, but also the systematic deviations demonstrated in Fig.~\ref{fig:NoCor} have been nearly completely removed over a wide range of revolution times.     

\begin{figure}[htbp]
	\centering
	\includegraphics[scale=0.42]{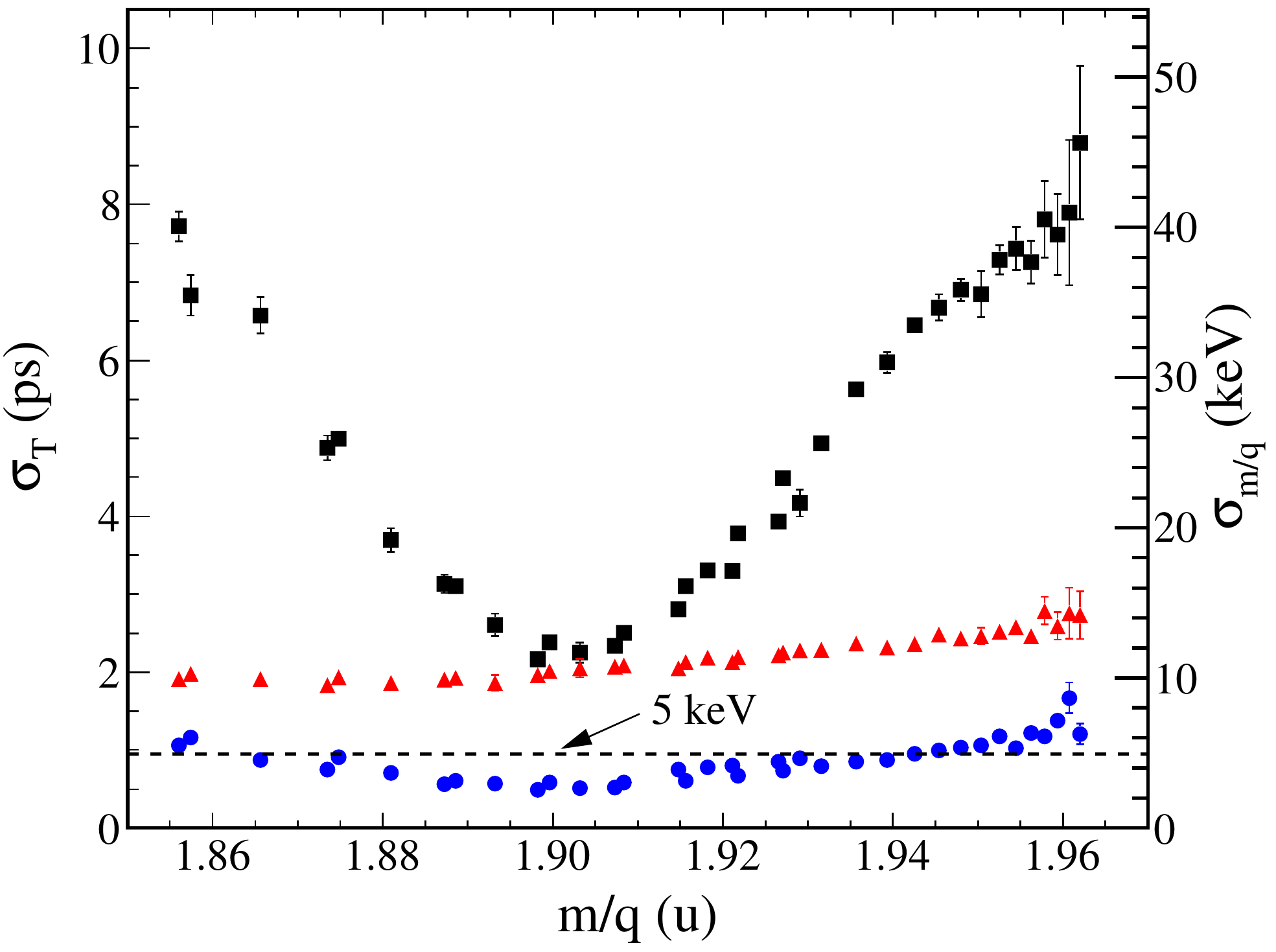}
	\caption{Standard deviations of the time peaks (left scale) extracted from original revolution time spectrum (filled black squares), and from  newly-constructed $T_{\rm fix}$ spectrum with (blue circles) and without (red triangles) magnetic field drift correction. The absolute accuracies of mass-to-charge ratios are given on the right scale.
	}
	\label{fig:st_m}
\end{figure}

To demonstrate the power of the $B\rho$-defined IMS in comparison with the conventional IMS, 
the standard deviations of the TOF peaks derived from the $T_{\rm exp}$ and $T_{\rm fix}$ spectra are shown in Fig.~\ref{fig:st_m}. 
The standard deviations of the TOF peaks in the original TOF spectrum, $\sigma_T$, have a parabolic dependence versus $m/q$.
$\sigma_T$ approaches minimum at approximately 2 ps only for a limited number of nuclides (isochronicity window). 
In the re-constructed $T_{\rm fix}$ spectrum without the field drift correction, 
$\sigma_T \approx 2$~ps has been achieved for basically all nuclides in the entire $m/q$-range. 

We emphasize that this was done without reducing the $B\rho$-acceptance of neither the ring nor the transfer line.
Here the resolving power is limited by the magnetic field drifts. 
After the field drift correction, $\sigma_T=0.5$ ps is achieved in the isochronicity window, 
corresponding to the mass resolving power of $3.3\times 10^5$ (FWHM). 
The mass resolving power at the edges of the spectrum has been improved by a factor of about 8.
The right scale in Fig.~\ref{fig:st_m} shows the corresponding absolute mass precisions for $m/q$ values.
It is emphasized that the mass precision of merely $5\cdot q$~keV can be achieved for just a single stored ion.

\section{Mass results and discussions}\label{sec:mass}

\subsection{Re-determined masses}

\begin{table*}
	\caption{Experimental mass excesses, MEs, obtained in this work, from an earlier CSRe measurement~\cite{Zhang2018,Zhang-2012,Yan-2013,Shuai2014} and from the literature. Also the recent Penning-trap measurements for $^{44g,44m}$V~\cite{Puentes2020}, $^{52g,52m}$Co~\cite{Nesterenko2017}, $^{56}$Cu~\cite{Valverde2018}, $^{51}$Fe~\cite{Ong2018} and AME2016 for $^{43}$Ti~\cite{Huang2017} are included. 
	}\label{tab:MEvalue}
	\begin{ruledtabular}
		\begin{tabular}{ccccccc}
			Atom  & $N$ & \makecell[c]{ME (keV)\\This work}  & \makecell[c]{ME (keV)\\Earlier CSRe}  & $\Delta$ME (keV) 	& \makecell[c]{ME (keV)\\Literature}  & $\Delta$ME (keV) \\ 
			\hline
			$^{44g}$V      & 601  &    $-23800.4(7.1)$	&  $-23827(20)	$   & $-26(21)	$   & $-23804.9(8.0)$~\cite{Puentes2020}	&  $-4.5(11)$   \\
			$^{44m}$V      &      &    $-23534.3(7.3)$	&  $-23541(19)	$   & $-6(20)	$   & $-23537(5.5)$~\cite{Puentes2020}	    &  $-2.7(9.1)$\\
			$^{46}$Cr      & 745  &    $-29477.2(2.6)$	&  $-29471(11)	$   & $6(11)	$   &                       &  \\
			$^{48}$Mn      & 685  &    $-29290.4(2.9)$	&  $-29299(7)	$   & $-9(8)	$   &                       &  \\
			$^{50}$Fe      & 782  &    $-34475.8(2.9)$	&  $-34477(6)	$   & $-1(7)	$   &                       &  \\
			$^{52g}$Co     & 845  &    $-34352.6(5.5)$	&  $-34361(8)	$   & $-8(10)	$   & $-34331.6(6.6)$~\cite{Nesterenko2017}	&  $21(9)$\\
			$^{52m}$Co     &  	  &    $-33973(10.6)$	&  $-33974(10)	$   & $-2(15)	$   & $-33958(11)$~\cite{Nesterenko2017}	&  $15(15)$\\
			$^{54}$Ni      & 1254 &    $-39285.4(2.7)$	&  $-39278.3(4)	$   & $7(5)		$   &                       &  \\
			$^{56}$Cu      & 294  &    $-38622.6(6.0)$	&  $-38643(15)	$   & $-21(16)	$   & $-38626.7(7.1)$~\cite{Valverde2018}	&  $-3.9(9.3)$\\
			$\rm ^{41}Ti$  & 25   &    $-15724.3(18.7)$	 & $-15697.5(27.9) $    & $26.8(33.6)$  &  & \\
			$\rm ^{43}V	$  & 9    &    $-17899.3(31.6)$	 & $-17916.4(42.8) $    & $-17.0(53.2)$ &  & \\
			$\rm ^{45}Cr$  & 57   &    $-19474.6(11.0)$	 & $-19514.8(35.4) $    & $-40.2(37.1)$ &  & \\
			$\rm ^{47}Mn$  & 18   &    $-22560.8(19.2)$	 & $-22566.4(31.7) $    & $-5.6(37.0)$  &  & \\
			$\rm ^{49}Fe$  & 86   &    $-24671.6(8.4)$	 & $-24750.7(24.2) $    & $-79.1(25.6)$ &  & \\
			$\rm ^{51}Co$  & 48   &    $-27386.0(11.5)$   & $-27342.1(48.4) $    & $43.8(49.8)$  &  & \\
			$\rm ^{53}Ni$  & 168  &    $-29613.7(5.8)$	 & $-29630.8(25.2) $    & $-17.1(25.8)$ &  & \\
			$\rm ^{55}Cu$  & 11   &    $-31807.0(24.8)$   & $-31635.4(155.6)$    & $171.6(157.5)$  &  & \\
			$^{43}$Ti      & 757  &    $ -29302.2(4.2)$	&  $-29306(9)$	& $-4(10)$	& $-29321(7)$~\cite{Huang2017}	    &  $19(8)$\\
			$^{51}$Fe      & 108  &    $ -40201.9(15.9)$	&  $-40198(14)$	& $4(21)$   & $-40189.2(1.4)$~\cite{Ong2018}	&  $13(16)$\\
		\end{tabular}
	\end{ruledtabular}
\end{table*}

\begin{figure}[htb]
	\centering
	\includegraphics[scale=0.40]{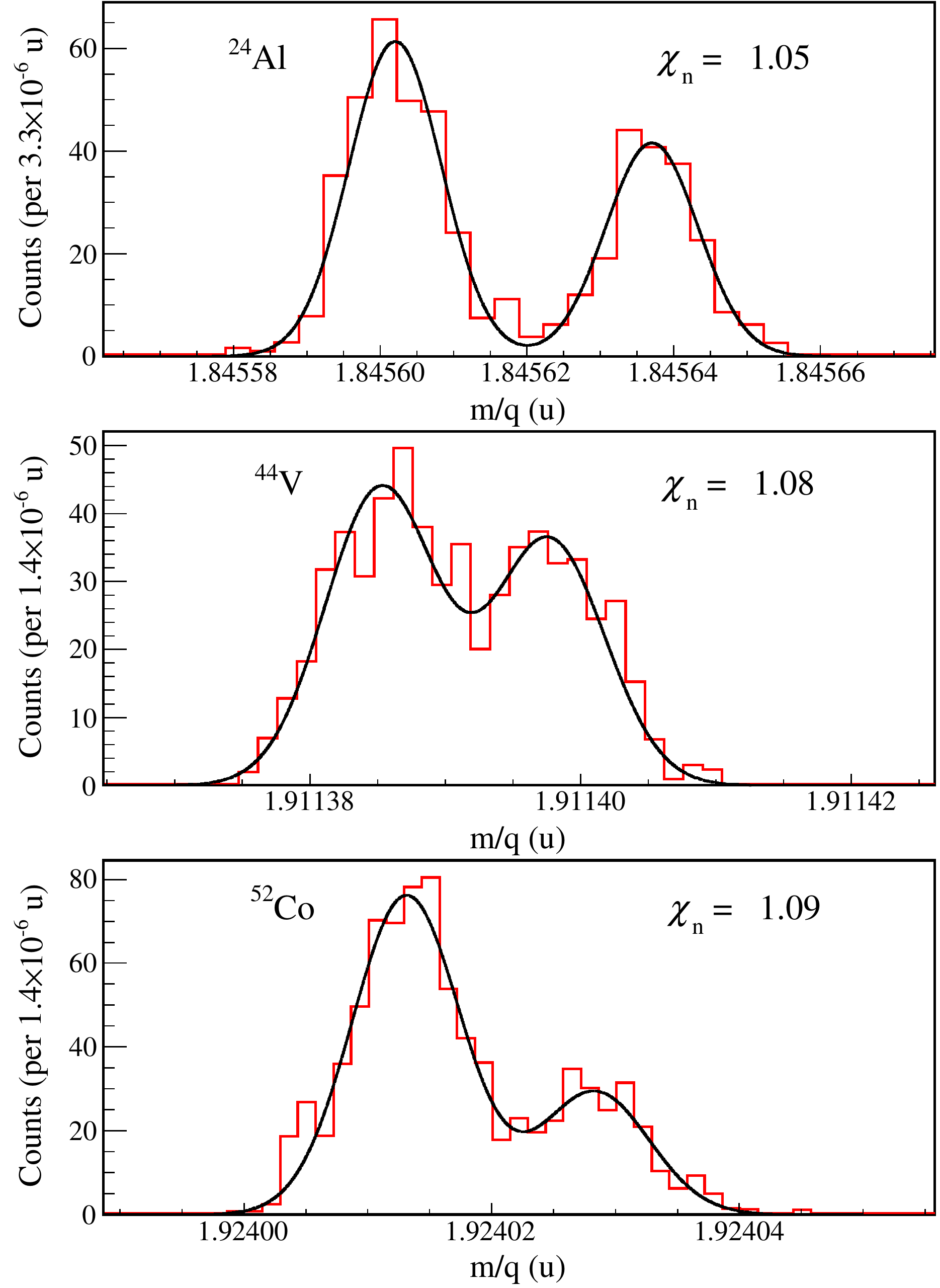}
	\caption{$m/q$ spectra of $^{24,24m}$Al, $^{44,44m}$V, and $^{52,52m}$Co obtained from the $B\rho$-defined IMS. Solid lines are fit results using two Gaussian functions of identical standard deviation.}
	\label{fig:mvqspec}
\end{figure}

Each individual $m/q$ value and the uncertainty were obtained in the event-by-event analysis. The $m/q$ values were put into a histogram forming the $m/q$ spectrum.
For the ion species with well-separated peaks in the $m/q$ spectrum, the weighted-average $m/q$ was directly derived and converted~\cite{Zhang2018} into atomic mass excess. For the nuclides with a low-lying isomer, two peaks are overlapped such as the cases of $^{24,24m}$Al, $^{44,44m}$V, and $^{52,52m}$Co. Therefore, the mean $m/q$ values were deduced by fitting the $m/q$ spectrum using two Gaussian functions with an identical standard deviation. The fit results are given in Fig.~\ref{fig:mvqspec}. The normalized $\chi_n$ values are all close to unity, indicating that the peak shapes can be well described by the proposed fitting functions.   

Table~\ref{tab:MEvalue} presents the newly determined MEs for the nuclides that were not used in the precedures of $B\rho(C)$ construction and field drift correction. Therefore, these MEs should be considered to be from independent measurements. The re-determined masses for the $T_z=-1$ nuclides have been reported in Ref.~\cite{PRL-part} emphasizing that the $B\rho$-defined IMS can provide masses of neutron deficient nuclides with the precisions comparable to those from the conventional Penning-trap mass spectrometer. 

The mass excess of $^{43}$Ti obtained in a previous CSRe experiment 
utilizing a single TOF detector~\cite{Zhang2018} was 15~keV larger than the AME2016 value~\cite{Huang2017}. 
This small deviation was attributed to a possible mixture with a low-lying isomer at $E_x=313$~keV. 
However, this isomer has a half-life of only 11.9 ${\rm \mu}$s~\cite{Kondev2021}, which can hardly affect our mass measurements. 
Our new results confirm the previous CSRe result, with the mass being 19 keV larger than the AME2016 value.
The latter was determined from two energy measurements in transfer reactions~\cite{ALDRIDGE1967,Mueller1977}. 
Our IMS measurement is the first direct mass measurements for this nuclide.
\begin{figure}[htb]
	\centering
	\includegraphics[scale=0.3]{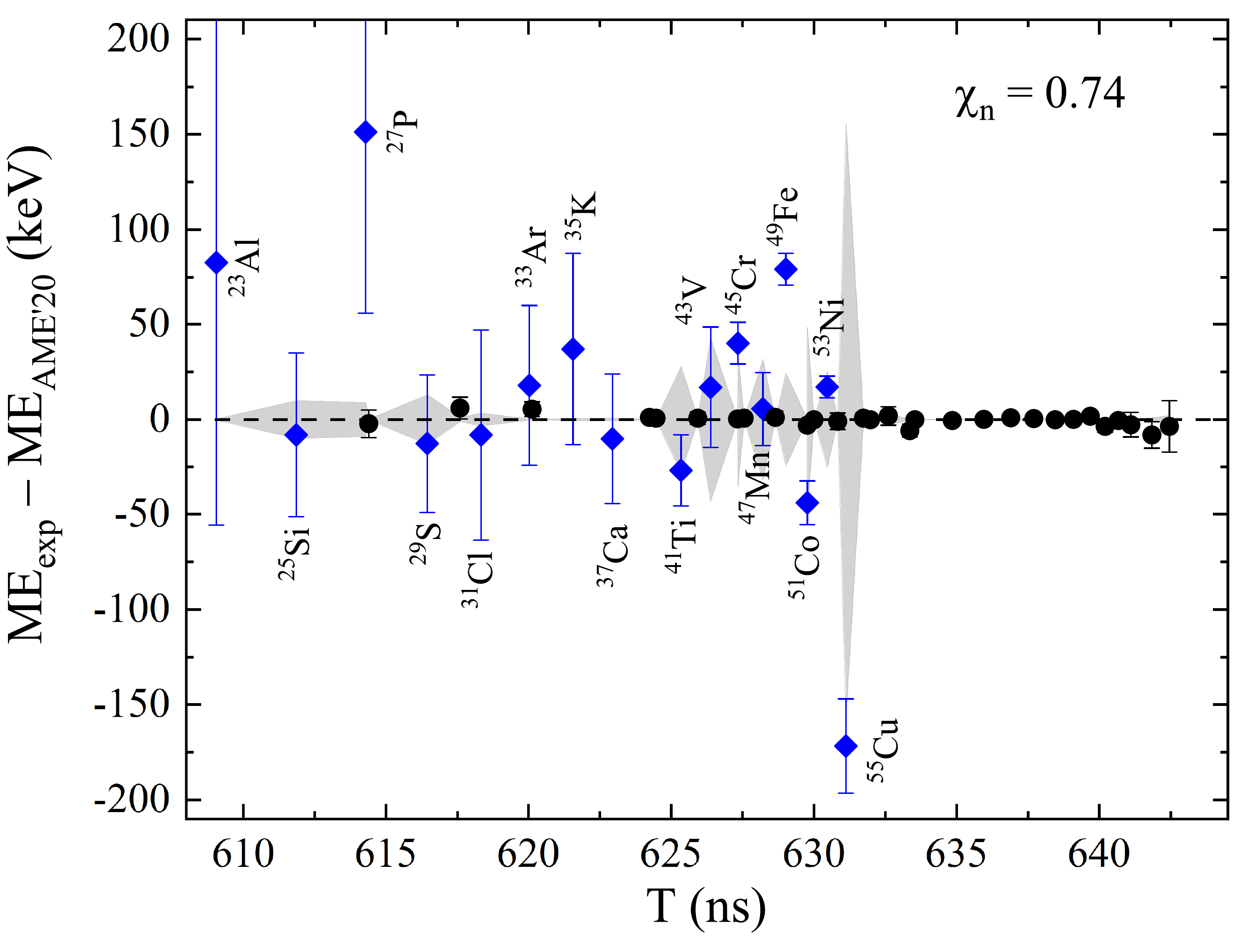}
	\caption{Comparison of the re-determined masses with literature values. 
		The grey shadow represents the mass uncertainties in the latest atomic mass evaluation~\cite{Wang2021}.
		The blue diamonds indicate the nuclides not used in the correction procedure of the magnetic field drifts. }
	\label{fig:dmeresult2}
\end{figure}

A comparison of newly-determined masses with literature values is shown in Fig.~\ref{fig:dmeresult2}. The filled black circles represent the nuclides that were used in constructing the $B\rho(C)$ function and in the field drift correction. Although they are the references in the mass determination, their values can be re-determined.
For this purpose, each of them, one by one, is assumed to be unknown and is calculated from the remaining reference masses.
The normalized $\chi_n=0.74$ for the re-determined reference masses indicates that the quoted errors are conservative and no additional systematic errors need to be considered.

The re-determined masses for the $T_z=-3/2$ nuclides are compared with literature ones in Fig.~\ref{fig:dmeresult2}. 
Except for $^{49}$Fe, the present mass excesses are in good agreement with the previously-measured values by using the conventional IMS~\cite{Zhang-2012,Yan-2013,Shuai2014}. 
The mass accuracies are significantly increased allowing us to investigate the mirror symmetry of empirical residual proton-neutron interactions ($p$-$n$ interaction), as well as to test the  famous isobaric multiplet mass equation (IMME). 

\subsection{Mirror symmetry of residual $p$-$n$ interactions}

The binding energy of a nucleus $B(Z,N)$, derived directly from atomic masses, embodies the sum of overall interactions inside the nucleus. 
Differences of masses or binding energies isolate specific nucleonic interactions and provide signatures and different features of the  nuclear structures. 
For example, the nucleon pairing correlations are clearly exhibited by the zig-zag pattern of one-nucleon separation energies with changing $N$ and $Z$, and the shell closures can be revealed by kinks in smooth trends of two-nucleon separation energies. 
Apart from the one- and two-nucleon separation energies, a double difference of masses or binding energies, denoted as $\delta V_{pn}$, has been used to isolate the average interaction strength between the last proton and the last neutron. 
Conventionally, $\delta V_{pn}$ values are derived 
according to~\cite{PhysRevLett.74.4607}
\begin{equation}\label{eq:dv1}
\begin{aligned}
\delta V_{pn}^{ee}(Z,N)=&\frac{1}{4}[B(Z,N)-B(Z,N-2) \\
&-B(Z-2,N)+B(Z-2,N-2)],
\end{aligned}
\end{equation} 
\begin{equation}\label{eq:dv2}
\begin{aligned}
\delta V_{pn}^{oe}(Z,N)=&\frac{1}{2}[B(Z,N)-B(Z,N-2) \\
&-B(Z-1,N)+B(Z-1,N-2)],
\end{aligned}
\end{equation}
\begin{equation}\label{eq:dv3}
\begin{aligned}
\delta V_{pn}^{eo}(Z,N)=&\frac{1}{2}[B(Z,N)-B(Z,N-1) \\
&-B(Z-2,N)+B(Z-2,N-1)],
\end{aligned}
\end{equation}
\begin{equation}\label{eq:dv4}
\begin{aligned}
\delta V_{pn}^{oo}(Z,N)=&[B(Z,N)-B(Z,N-1)- \\
&B(Z-1,N)+B(Z-1,N-1)],
\end{aligned}
\end{equation}
and termed as the empirical residual proton-neutron interactions. 
Eqs.~(\ref{eq:dv1}) through (\ref{eq:dv4}) are suited for nuclei with $(Z,N)=$ even-even, odd-even, even-odd, and odd-odd, respectively. 

It has been recognized that the $p$-$n$ interactions are closely related to many nuclear structure phenomena such
as the onset of collectivity and deformation~\cite{RevModPhys.34.704,PhysRevLett.102.082501}, changes
of underlying shell structure~\cite{HEYDE1985303}, and phase transitions in nuclei~\cite{HEYDE1985303,FEDERMAN1977385,FEDERMAN197829}. 
From systematic investigations of $\delta V_{pn}$ throughout the mass surface, strong orbital dependence of $\delta V_{pn}$ has been revealed for the self-conjugate nuclei and in the regions of doubly shell closures~\cite{PhysRevLett.94.092501,PhysRevC.73.034315}. 

Inspecting the available $\delta V_{pn}$ data of the nuclides on both sides of the $Z=N$ line, one observed a mirror symmetry of the residual $p$-$n$ interactions~\cite{Zhang2018}. 
Indeed, such a mirror symmetry was pointed out many years ago by J\"anecke~\cite{PhysRevC.6.467}, and recently was used by Zong {\it et al.}~\cite{PhysRevC.102.024302} to make high-precision mass predictions for the very neutron-deficient nuclei. 
\begin{figure}[htb]
	\centering
	\includegraphics[scale=0.40]{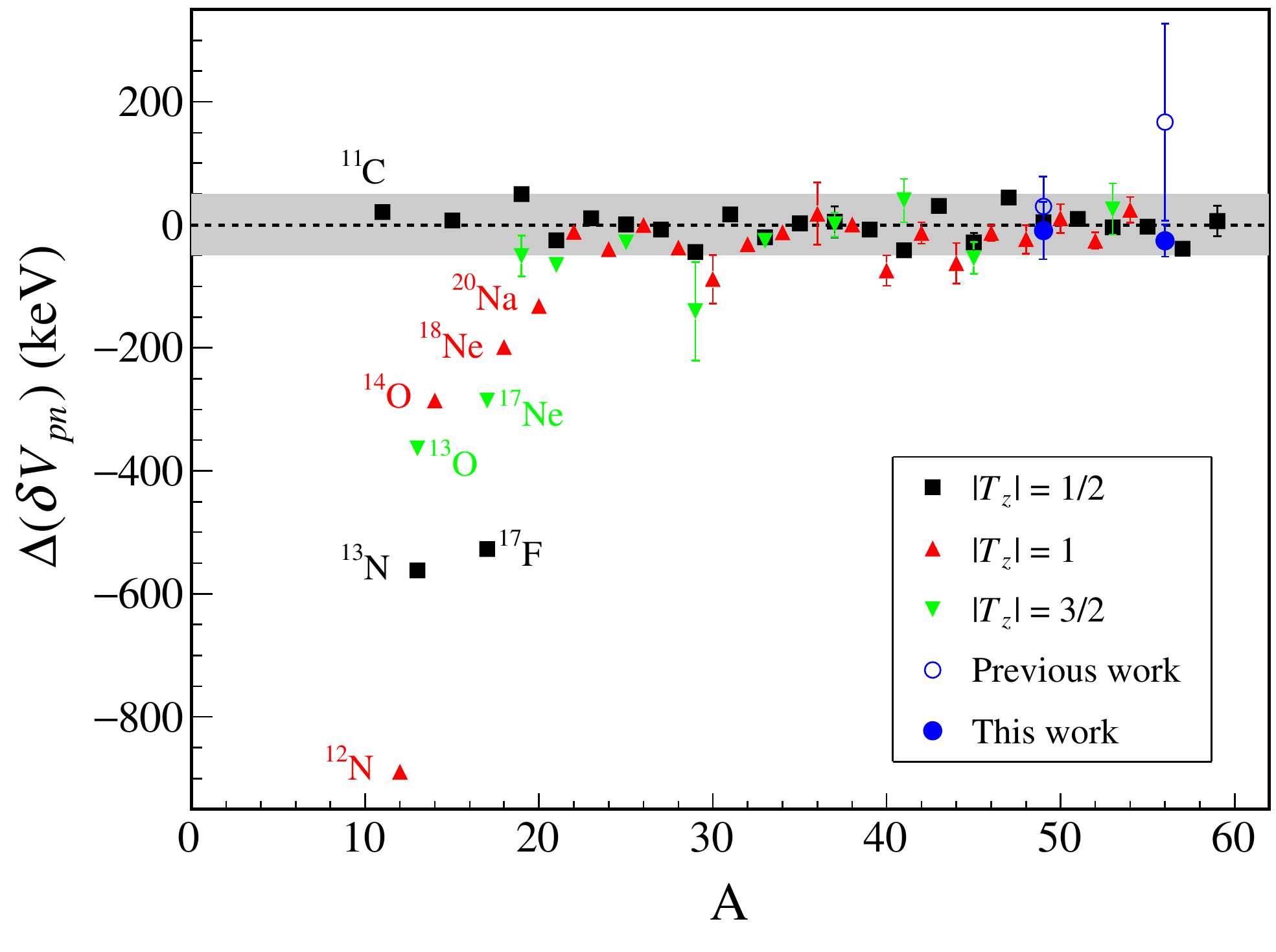}
	\caption{Differences of $\delta V_{pn}$ of mirror nuclei. The grey shadow indicates an error band of 50 keV. The neutron-deficient partners are indicated for the mirror pairs with $A\le 20$. }
	\label{fig:mirror}
\end{figure}

We present in Fig.~\ref{fig:mirror} the energy differences, defined as $\Delta(\delta V_{pn})=\delta V_{pn}(T_z^{<},A)-\delta V_{pn}(T_z^{>},A)$. Here $T_z^{<}$/$T_z^{>}$ represents the negative/positive value of $T_z$ of the mirror nuclei. 
The $\delta V_{pn}$ values are calculated according to Eqs.~(\ref{eq:dv1}) through (\ref{eq:dv4}) using the mass data in Ref.~\cite{Wang2021} and the new masses in Table~\ref{tab:MEvalue}. 
It is seen that $\Delta(\delta V_{pn})$ values scatter around zero within an error band of $\pm 50$~keV for the $A>20$ mirror pairs, indicating that the mirror symmetry of $\delta V_{pn}$ holds well in this mass region. It is worthwhile to note that the mass of $^{55}$Cu is re-determined with significantly-improved precision, and its value is 172~keV more bound than the previous one~\cite{Yan-2013}. Using this new mass, the calculated $\delta V_{pn}$ of $^{56}$Cu is approximately equal to that of $^{56}$Co. Consequently, $\Delta(\delta V_{pn})=\delta V_{pn}(^{56}{\rm Cu})-\delta V_{pn}(^{56}{\rm Co})$ fits well into the general systematics at a high level of accuracy (see Fig.~\ref{fig:mirror}). 
In turn this result provides a further confirmation of the reliability of the new mass value of $^{55}$Cu.

In the lighter mass region with $A\le 20$, the so-called mirror symmetry of $\delta V_{pn}$ is apparently broken in a few mirror pairs (see Fig.~\ref{fig:mirror}), i.e., the $\delta V_{pn}$ values of the neutron-deficient nuclei are systematically smaller than those of the corresponding neutron-rich partners. Such a mirror-symmetry breaking was pointed out and is simply attributed to a binding energy effect~\cite{PhysRevC.6.467}. 

It is noted that the calculations of $\delta V_{pn}$ require four binding energy values, and one of the four nuclei is particle\YAL{-}unbound in the cases of mirror-symmetry breaking. The negative $\Delta(\delta V_{pn})$ values shown in Fig.~\ref{fig:mirror} could be understood, at least qualitatively, by considering a proton halo-like structures of these very neutron-deficient nuclei. 

Taking $^{13}$N as an example, its empirical $p$-$n$ interaction is calculated by
\begin{equation}\label{eq:dv5}
\begin{aligned}
\delta V_{pn}(^{13}{\rm N)}=\frac{1}{2}[B(^{13}{\rm N})-B(^{11}{\rm N})-B(^{12}{\rm C})+B(^{10}{\rm C})].
\end{aligned}
\end{equation}
Among the four nuclei, only $^{11}$N is proton unbound with $S_p=-1378$ keV~\cite{PhysRevC.100.024306}. The valance proton in $^{11}$N occupies the unbound $p_{1/2}$ or $s_{1/2}$ single particle orbit, and the wave function or density distribution would be spatially expanded, showing most probably a halo-like structure and leading to a reduction of the Coulomb energy~\cite{PhysRevC.100.064303}. 
Since the Coulomb interaction is repulsive, the reduction of Coulomb energy due to the halo-like structure will lead to an increase of total binding energy with respect to the $normal$ structure. The final consequence of the halo-like structure of $^{11}$N would be the decreasing of $\delta V_{pn}(^{13}{\rm N)}$, see Eq.~(\ref{eq:dv5}), giving a negative value of $\Delta(\delta V_{pn})$ as shown in Fig.~\ref{fig:mirror}.

\subsection{Validity of isospin multiplet mass equation}

The masses of a set of isobaric analog states (IASs) can be described by the famous quadratic isospin multiplet mass equation (IMME)~\cite{RevModPhys.51.527},
\begin{equation}\label{IMME_equ}
{\rm ME}(A,T,T_{z}) =a(A,T)+b(A,T)T_{z}+c(A,T)T_{z}^{2},
\end{equation}
where MEs are mass excesses of IASs of a multiplet with fixed mass number $A$ and total isospin $T$. $T$ is equal to or is larger than the projection of $T$, $T_z=(N-Z)/2$, for a specific nucleus. The coefficients $a, b$, and $c$ depend on $A,~T$, and other quantum
numbers such as the spin-and-parity $J^{\pi}$, but are independent of $T_z$. 
The quadratic form of the IMME, i.e., Eq.~(\ref{IMME_equ}),
is commonly considered to be accurate within uncertainties of a few tens of keV. In this context, precision mass
measurements can be used for testing its validity (see Ref.~\cite{Zhang-2012} and references therein). 
Typically one adds to Eq.~(\ref{IMME_equ}) extra terms such as $dT_z^3$ or/and $eT_z^4$, which provide a measure of the breakdown of the quadratic form of the IMME. 
Numerous measurements have been performed investigating the validity of the IMME. Reviews and compilations of existing data can be found in
Refs.~\cite{MACCORMICK201461,LAM2013680} and references cited therein. 

\begin{table*}[!htb]
	\caption{ME values of $T=3/2$, $A=55$ isobaric quartet. The data with symbol ``\#" are from AME2020~\cite{Wang2021}, and the ME value of $^{55}$Cu is from this work. The excitation energies of the IASs in $^{55}$Ni and $^{55}$Co are shifted by $-5.5$ keV and $10.3$ keV, respectively. See text for details.}
	\label{tab:IMME}
	\centering
	\begin{ruledtabular}
		\begin{tabular}{lccll}
			Atom               &  $T_z$   & ME(g.s.) (keV)  &  $E_x$ (keV) &  ME(IAS) (keV)\\
			\hline
			$^{55}{\rm Fe}$    & $3/2$  & $-57481.4(3)^{\#}$ & $0$         & $-57481.4(3)$    \\
			$^{55}{\rm Co}$    & $1/2$  & $-54030.0(4)^{\#}$ & $4731.8(40)$    & $-49298.2(40)$      \\
			$^{55}{\rm Ni}$    & $-1/2$ & $-45336.0(7)^{\#}$ & $4593.5(14)$    & $-40742.5(16)$      \\
			$^{55}{\rm Cu}$    & $-3/2$ & $-31807(25)$  & $0$             & $-31807(25)$   \\
			Quadratic fit: $\chi_n$ = 0.26 & &$a=-45067.2(28)$~keV & $b=-8556.2(38)$~keV & $c=186.7(36)$~keV \\
			Cubic fit: $d$ = $-1.2(47)$~keV & &$a=-45067.4(29)$~keV & $b=-8555.4(50)$~keV & $c=188.1(63)$~keV \\
		\end{tabular}
	\end{ruledtabular}
\end{table*}

The re-determined ME of $^{55}$Cu is 172 keV smaller than the previous value~\cite{Yan-2013}. Using this new mass, four experimental masses of the  $T=3/2$ IASs are complete, and thus the validity of the quadratic form of the IMME can be tested reaching to the heaviest $A=55$ isospin quartet. 
Table~\ref{tab:IMME} collects the relevant data for the $A=55$, $T=3/2$ isospin quartet. 

It is noted that the $T=3/2$ IASs in $^{55}$Ni $(T_z=-1/2)$ and in $^{55}$Co $(T_z=+1/2)$ have strong isospin mixing with the close-lying $T=1/2$ states~\cite{PhysRevLett.111.262501,PhysRevC.17.1588}, leading to energy shifts for the pure IASs.
Their isospin mixing matrix elements were inferred from experiments to be 9(1) keV and 13(4) keV~\cite{PhysRevLett.111.262501}, respectively. Consequently, the energy splitting between the two unperturbed $T=3/2$ and $T=1/2$ levels have been extracted according to Ref.~\cite{PhysRevLett.111.262501}, and the excitation energies of the corresponding IASs in NUBASE2020~\cite{Kondev2021} have been shifted by $-5.5$ keV and $10.3$ keV, respectively. The modified values are listed in Table~\ref{tab:IMME} for further analysis. 

\begin{figure}[htb]
	\centering
	\includegraphics[scale=0.29]{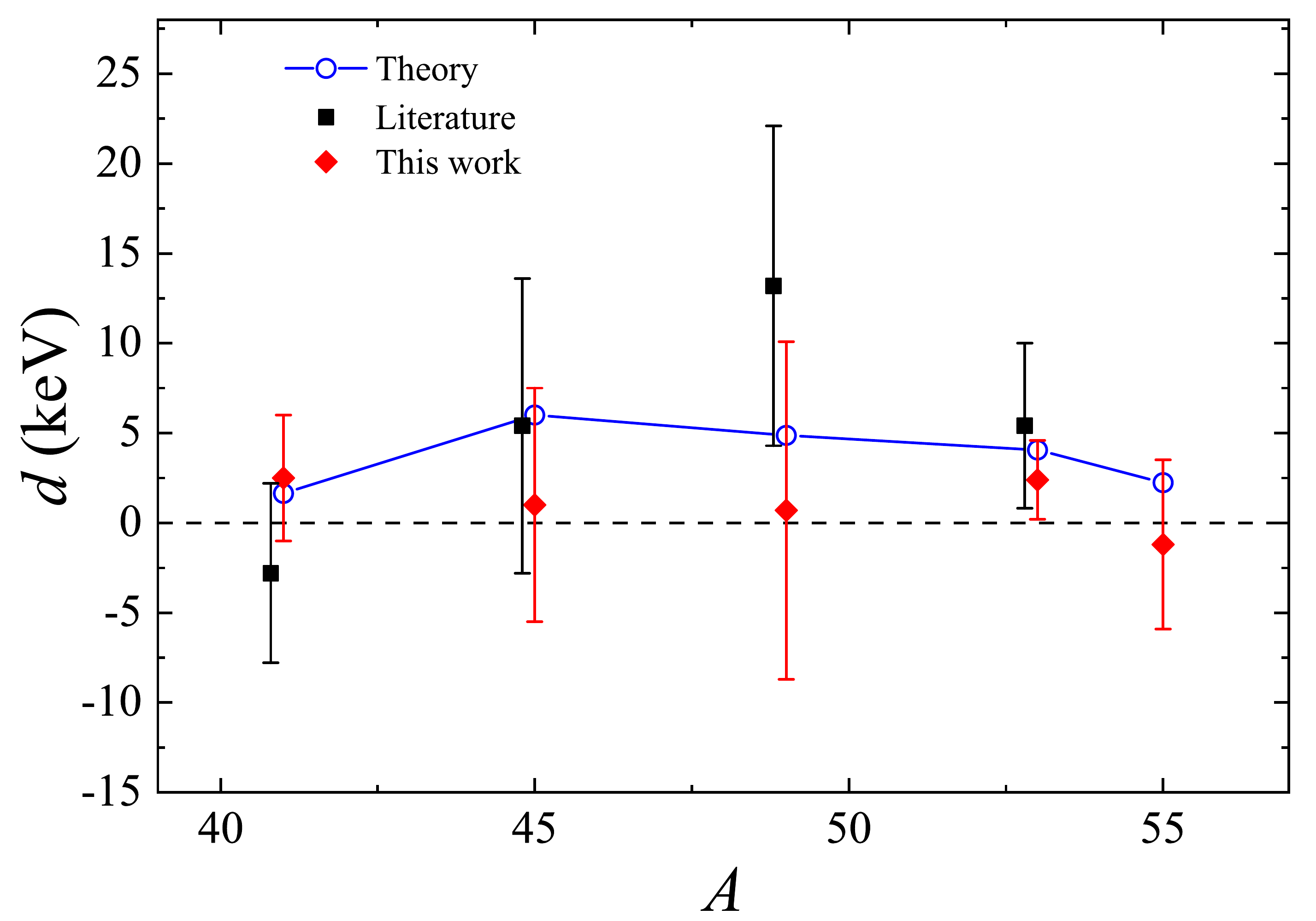}
	\caption{$d$ coefficients of the cubic form of IMME for the $T=3/2$, $A=41,~45,~49,~53$, and $55$ isospin quartets. The solid line connects the predicted $d$ values from theoretical calculations~\cite{PhysRevC.99.014319}.}  
	\label{fig:d-coef}
\end{figure} 

The mass data were fitted using Eq.~(\ref{IMME_equ}) with the normalized $\chi_n$ = 0.26. The obtained $a,~b$, and $c$ coefficients are given in Table~\ref{tab:IMME} which are consistent with systematics in Refs.~\cite{MACCORMICK201461,LAM2013680}. 
In addition, the mass data were also fitted by the cubic form of IMME by adding the $dT_z^3$ term to Eq.~(\ref{IMME_equ}). The obtained $d$ coefficient is $-1.2(47)$ keV and is compatible with zero. 
These results indicate that Eq.~(\ref{IMME_equ}) can well describe the mass data, i.e., the quadratic form of the IMME holds well for the $A=55$, $T=3/2$ isospin quartet.      

Similar procedure has been applied to the $T=3/2$, $A=41,~45,~49$, and $53$ isospin quartets using the new mass data in Table~\ref{tab:MEvalue}. 
The obtained $d$ coefficients are shown in Fig.~\ref{fig:d-coef}. 
Comparing with the previous results of Fig.~2 in Ref.~\cite{Zhang-2012}, one concludes that the trend of a gradual increase of $d$ with $A$ in the $fp$ shell~\cite{Zhang-2012} is not confirmed, at least at the present level of accuracy. 
Given the fact that all extracted $d$ coefficients are compatible with zero, the quadratic form of the IMME is valid for the cases investigated here. 
Figure~\ref{fig:d-coef} also shows the $d$ values from theoretical calculations~\cite{PhysRevC.99.014319}. 
The predicted non-zero $d$ coefficients for these $T=3/2$ isospin quartets can not be 
yet ruled out due to large experimental uncertainties. More precise mass measurements for the associated IASs are still needed to draw a definite conclusion. 

\section{Summary and Outlook}\label{sec:conclusion}

An improved isochronous mass spectrometry, the $B\rho$-defined IMS, has been developed at the experimental cooler-storage ring CSRe.
The measurements of both the revolution time and the velocity of every stored 
ion enable us to construct the $B\rho(C)$ function, which is a universal mass calibration curve for all simultaneously stored ions.
The method for constructing the $B\rho(C)$ function is described in detail by using the experimental data in the mass measurements of $^{58}$Ni projectile fragments.
The high mass resolving power over the whole $B\rho$-acceptance of the storage ring is realized.
The uncertainty band as small as $\approx5\cdot q$~keV is achieved,
indicating that a single short-lived ($T_{1/2}\gtrsim100~\mu$s) ion is now sufficient for its mass determination with $\approx5\cdot q$~keV precision.

Masses of several $T_z=-3/2$ $fp$-shell nuclides were re-determined with high accuracy. The precision mass value of $^{55}$Cu complete the four masses of the $A=55,$ $T=3/2$ isospin quartet, and thus the isospin multiplet mass equation have been validated up to the heaviest quartet with $A=55$. 
The new masses are also used to investigate the mirror symmetry of the empirical residual $p$-$n$ interactions, which could be employed, in turn, to test the reliability of newly-measured masses, or to predict the masses of very neutron-deficient nuclei. 
It is also pointed out that the mirror-symmetry breaking of $\delta V_{pn}$ in the $A\le 20$ region may be due to the halo-like structure of a proton-unbound nucleus involved in the extraction of $\delta V_{pn}$. 
Further experimental as well as theoretical studies are needed to address this issue. 

The merits of the novel IMS are demonstrated by the dramatically increased sensitivity and accuracy of the measurements. Next-generation storage rings~\cite{STECK2020103811} are under construction, including the Spectrometer Ring (SRing)~\cite{WU2018} at the high-intensity Heavy-Ion Accelerator Facility (HIAF)~\cite{YANG2013,Liu2018} in China, and the Collector Ring (CR)~\cite{DOLINSKII2008,DOLINSKII2007} at the Facility for Antiproton and Ion Research (FAIR)~\cite{Litvinov2010,WALKER2013,Bosch-2013,Durante-2019} in Germany. Mass measurements of exotic nuclei are among the research programs of these storage rings~\cite{YAMAGUCHI2021103882}. Owing to the high resolving power, high accuracy, ultimate sensitivity, short measurement time, broadband, and background-free characteristics, the $B\rho$-defined IMS is the technique of choice for the future storage-ring based mass spectrometers. 
A further improvement of the $B\rho$-defined IMS is to work on the ion-optical setting of the ring to achieve a constant $\gamma_t$ over the full momentum acceptance, which will make the $B\rho(C)$ function well defined with a simple expression.

\begin{acknowledgments}
	The authors thank the staff of the accelerator division of IMP for providing stable beam. 
	This work is supported in part by the National Key R$\&$D Program of China (Grant No. 2018YFA0404401), 
	the Strategic Priority Research Program of Chinese Academy of Sciences (Grant No. XDB34000000), 
	CAS Project for Young Scientists in Basic Research (Grant No. YSBR-002),
	and the NSFC (Grants No 12135017, No. 12121005, No. 11961141004, No. 11905259, No. 11905261, No. 11975280). 
	Y.M.X. and C.Y.F. acknowledge the support from CAS ``Light of West China" Program. 
	Y.A.L. and R.S.S. are supported by the European Research Council (ERC) under the EU Horizon 2020 research and innovation programme (ERC-CG 682841 ``ASTRUm"). 
	T.Y. and S.S. were supported in part by JSPS and NSFC under the ``Japan-China Scientific Cooperation Program".
\end{acknowledgments}



%

\appendix
\section{}\label{app:pars}

In the present $B\rho$-defined IMS, Eq.~(\ref{eq:mvq}) is used for the mass determination which needs precise and accurate $B\rho$ and $v$ values of the ions of interest. Since $B\rho$ values are obtained from the $B\rho (C)$ functions, and the establishment of $B\rho (C)$ needs also the $v$ values of known-mass nuclides, it is of most importance to accurately determine the velocities according to Eq.~(\ref{eq:vel}). The $\Delta t^i_{\rm exp}$ values were determined using the method described in details in Ref.~\cite{Zhou2021}. 

\begin{figure}[htb]
	\centering
	\includegraphics[scale=0.3]{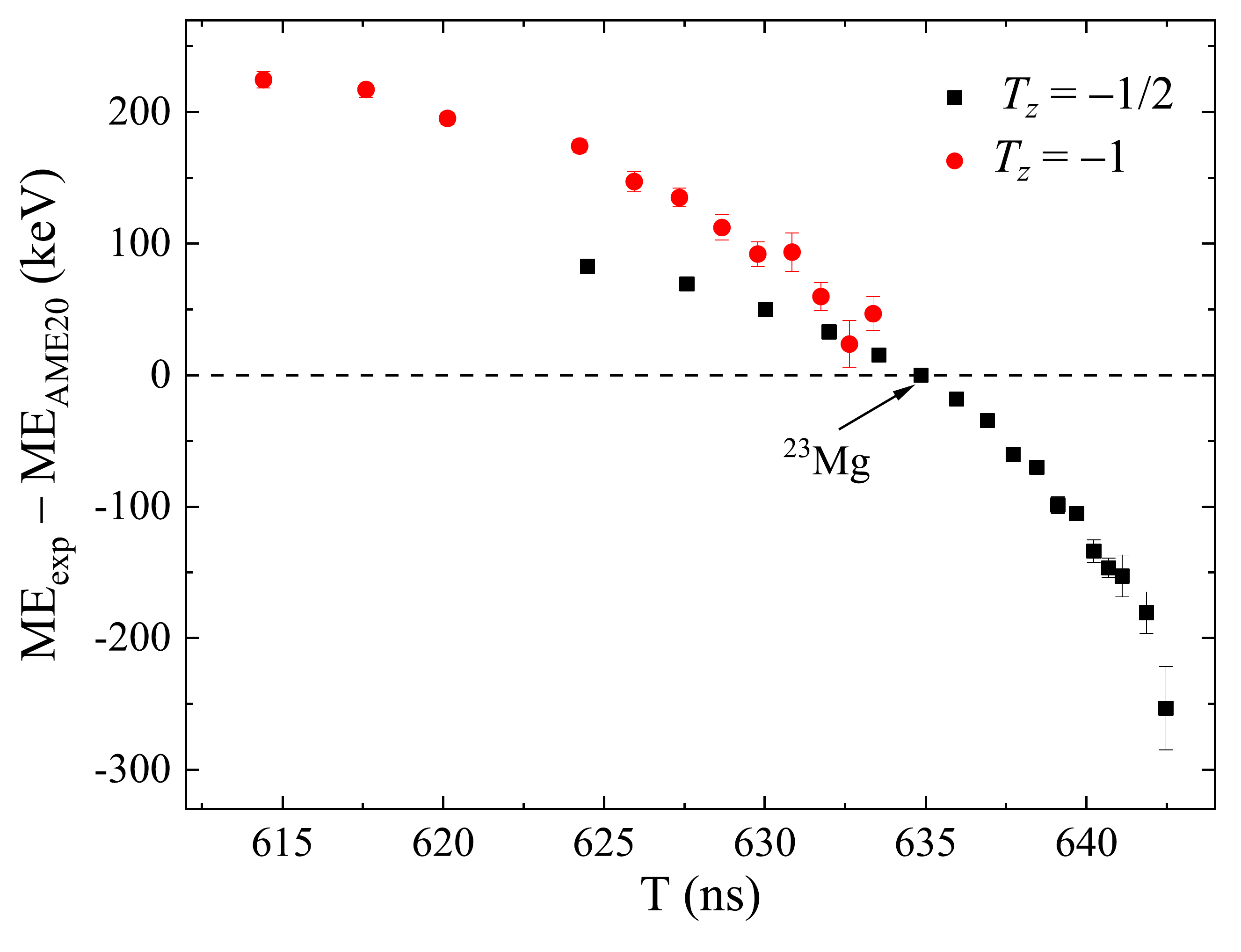}
	\caption{Comparison of the re-determined mass excess values ME$_{\rm exp}$ to the available values from the Atomic Mass Evaluation 2020 (AME2020)~\cite{Wang2021}.
		$^{23}{\rm Mg}^{12+}$ ions were used as a reference to establish the $B\rho=B\rho(C)$ function. 
		The divergence of the two $T_z$ series is due to different electric charges of the ions.}
	\label{fig:gDME_0}
\end{figure}

Using the initial $L$ and $\Delta t_{\rm d}$ values reported in Ref.~\cite{Yan2019} for velocity determinations, and the data points $\left\{(B\rho)_{\rm{exp}}^i,C_{\rm{exp}}^i\right\}$ of $^{23}$Mg for constructing the $B\rho(C)$ function (see Section~\ref{ssec:brhocfunc}), the $m/q$ values of all ions species have been obtained. 
The extracted mass excesses, ME$_{\rm exp}$, were compared to available data from the Atomic Mass Evaluation 2020 (AME2020)~\cite{Wang2021}, see Fig.~\ref{fig:gDME_0}. It is clearly seen that the re-determined masses deviate systematically from the literature ones. 
This is due to the biased $L$ and $\Delta t_{\rm d}$ values used in the velocity determination via Eq.~(\ref{eq:vel}) that can lead to different $B\rho(C)$ curves 
for different ion species and the systematic deviations will emerge in the mass determination.

\begin{figure*}[!htb]
	\centering
	\includegraphics[scale=0.6]{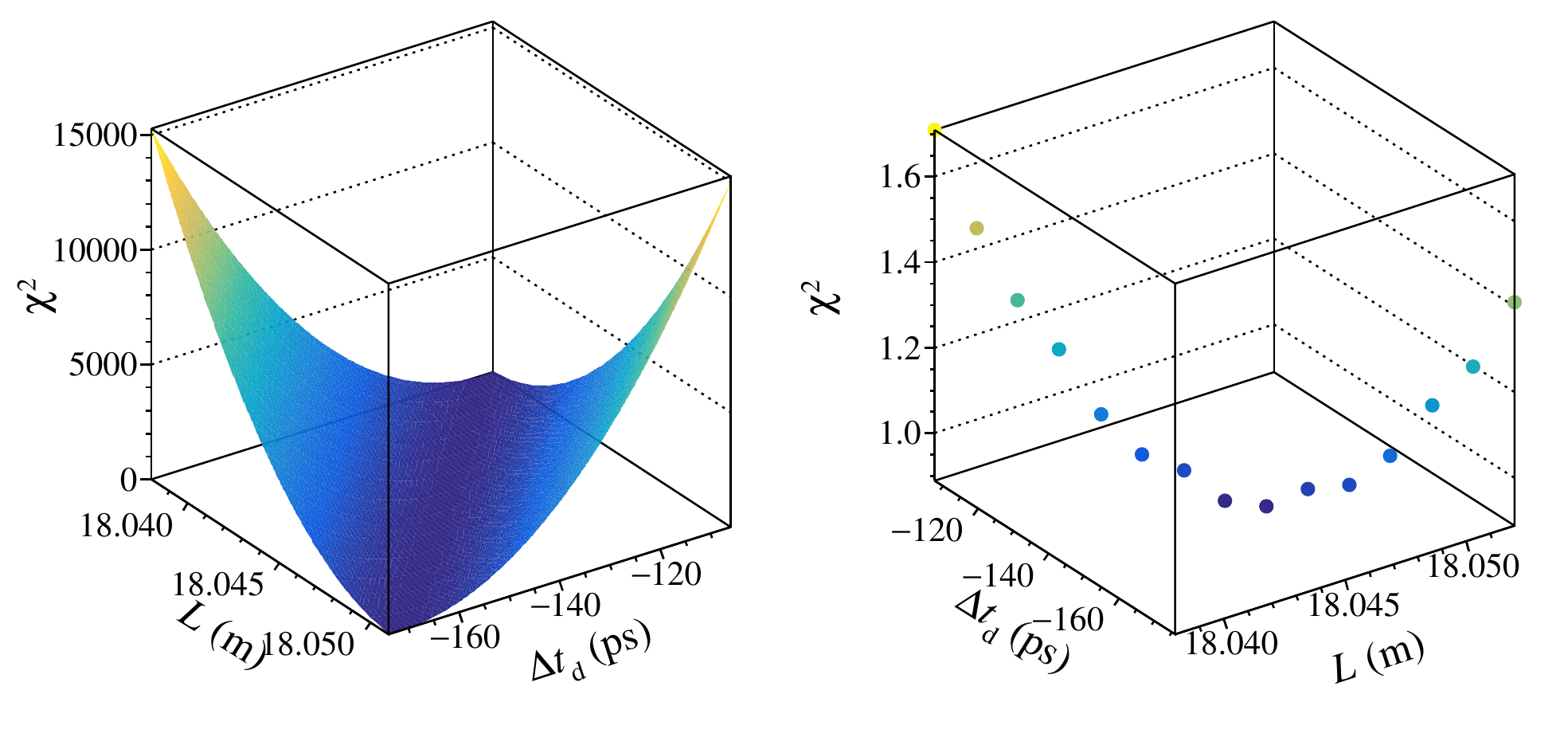}
	\caption{The $\chi^2$ obtained by varying $L$ and $\Delta t_{\rm d}$ values, (a) in the full parameter space and (b) around the minimum $\chi^2$.}
	\label{fig:ldtd}
\end{figure*}

To obtain the accurate $L$ and $\Delta t_{\rm d}$ values for velocity determinations, the $\chi^2$ defined as
\begin{equation}\label{eq:chi2}
\chi^2=\frac{1}{N_c}\sum_i{\frac{(ME_{\rm exp}^i-ME_{\rm AME}^i)^2}{\sigma_{\rm exp}^2+\sigma_{\rm AME}^2}}
\end{equation}
was minimized by varying $L$ and $\Delta t_{\rm d}$ for $N_c$ reference nuclides.
Ion species with tabulated mass uncertainties smaller than 5~keV~\cite{Wang2021} and with more than 100 counts recorded in the experiment were considered. 
Here the $ME_{\rm exp}^i$ and $ME_{\rm AME}^i$ indicate mass excesses of the nuclides re-determined in this work and from the literature~\cite{Wang2021}, respectively. The $\sigma_{\rm exp}$ and $\sigma_{\rm AME}$ are the corresponding mass uncertainties.

The $\chi^2$ obtained for various $L$ and $\Delta t_{\rm d}$ values is shown in Fig.~\ref{fig:ldtd}. 
It can be seen that these two parameters are highly correlated. 
The optimal values are $\Delta t_{\rm d}=-146.83$~ps and $L=18.046$~m. 
These two parameters are assumed to be constant for all ion species in a given experiment.

\end{document}